\documentclass[a4paper,fleqn]{cas-dc}
\usepackage[numbers]{natbib}
\def\tsc#1{\csdef{#1}{\textsc{\lowercase{#1}}\xspace}}
\tsc{WGM}
\tsc{QE}
\tsc{EP}
\tsc{PMS}
\tsc{BEC}
\tsc{DE}
\usepackage{picins}
\usepackage{subfig}%
\usepackage{graphicx}%
\usepackage{multirow}%
\usepackage{amsmath,amssymb,amsfonts}%
\usepackage{amsthm}%
\usepackage{mathrsfs}%
\usepackage[title]{appendix}%
\usepackage{xcolor}%
\usepackage{textcomp}%
\usepackage{manyfoot}%
\usepackage{booktabs}%
\usepackage{algorithm}%
\usepackage{algorithmicx}%
\usepackage{algpseudocode}%
\usepackage{listings}%
\usepackage{expl3} 
\PassOptionsToPackage{draft}{graphicx}
\begin{document}
	
\let\WriteBookmarks\relax
\def\floatpagepagefraction{1}
\def\textpagefraction{.001}
\shorttitle{Testing and Fault Tolerance Techniques for Carbon Nanotube-Based FPGAs}
\shortauthors{Siyuan Lu, Kangwei Xu, Peng Xie, Rui Wang, Yuanqing Cheng}

\title[mode=title]{Testing and Fault Tolerance Techniques for CNT-Based FPGAs}

\author[1, 4]{\textcolor{black}{Siyuan Lu}$^{1}$}[style=chinese]
\author[2]{\textcolor{black}{Kangwei Xu}$^{1}$}[style=chinese]
\author[1]{\textcolor{black}{Peng Xie}}[style=chinese]
\author[3]{\textcolor{black}{Rui Wang}}[style=chinese]
\author[1, 4]{\textcolor{black}{Yuanqing Cheng}\corref{cor1}}[style=chinese]

\affiliation[1]{organization={School of Integrated Circuit Science and Engineering},
                addressline={Beihang University},
                city={Beijing},
                postcode={100190},
                country={China}}

\affiliation[2]{organization={Department of Electronic Design Automation},
                addressline={Technical University of Munich (TUM)},
                city={Munich},
                postcode={80333},
                country={Germany}}

\affiliation[3]{organization={School of Computer Science and Engineering},
                addressline={Beihang University},
                city={Beijing},
                postcode={100190},
                country={China}}

\affiliation[4]{organization={Shenzhen Institute of Beihang University},
                city={Shenzhen},
                postcode={518000},
                country={China}}
                
\fntext[1]{Equal contribution.}                
\cortext[cor1]{Corresponding author.}

\begin{abstract}
As the semiconductor manufacturing process technology node shrinks into the nanometer-scale, the CMOS-based Field Programmable Gate Arrays (FPGAs) face big challenges in scalability of performance and power consumption. Multi-walled Carbon Nanotube (MWCNT) serves as a promising candidate for Cu interconnects thanks to the superior conductivity. Moreover, Carbon Nanotube Field Transistor (CNFET) also emerges as a prospective alternative to the conventional CMOS device because of high power efficiency and large noise margin. The combination of MWCNT and CNFET enables the promising CNT-based FPGAs. However, the MWCNT interconnects exhibit significant process variations due to immature fabrication process, leading to delay faults. Also, the non-ideal CNFET fabrication process may generate a few metallic CNTs (m-CNTs), rendering correlated faulty blocks. In this article, we propose a ring oscillator (RO) based testing technique to detect delay faults due to the process variation of MWCNT interconnects. Furthermore, we propose an effective testing technique for the carry chains in CLBs, and an improved circuit design based on the lookup table (LUT) is applied to speed up the fault testing of CNT-based FPGAs. In addition, we propose a testing algorithm to detect m-CNTs in CLBs.  Finally, we propose a redundant spare row sharing architecture to improve the yield of CNT-based FPGA further. Experimental results show that the test time for a 6-input LUT can be reduced by 35.49\% compared with conventional testing, and the proposed algorithm can achieve a high test coverage with little overhead. The proposed redundant architecture can repair the faulty segment effectively and efficiently.

\end{abstract}


\maketitle
\thispagestyle{empty}
\pagestyle{empty}

\section{Introduction}

\label{sec:intro}
\begin{figure}
\vspace{0.2cm}
	\centering
	\includegraphics[width=3.2in]{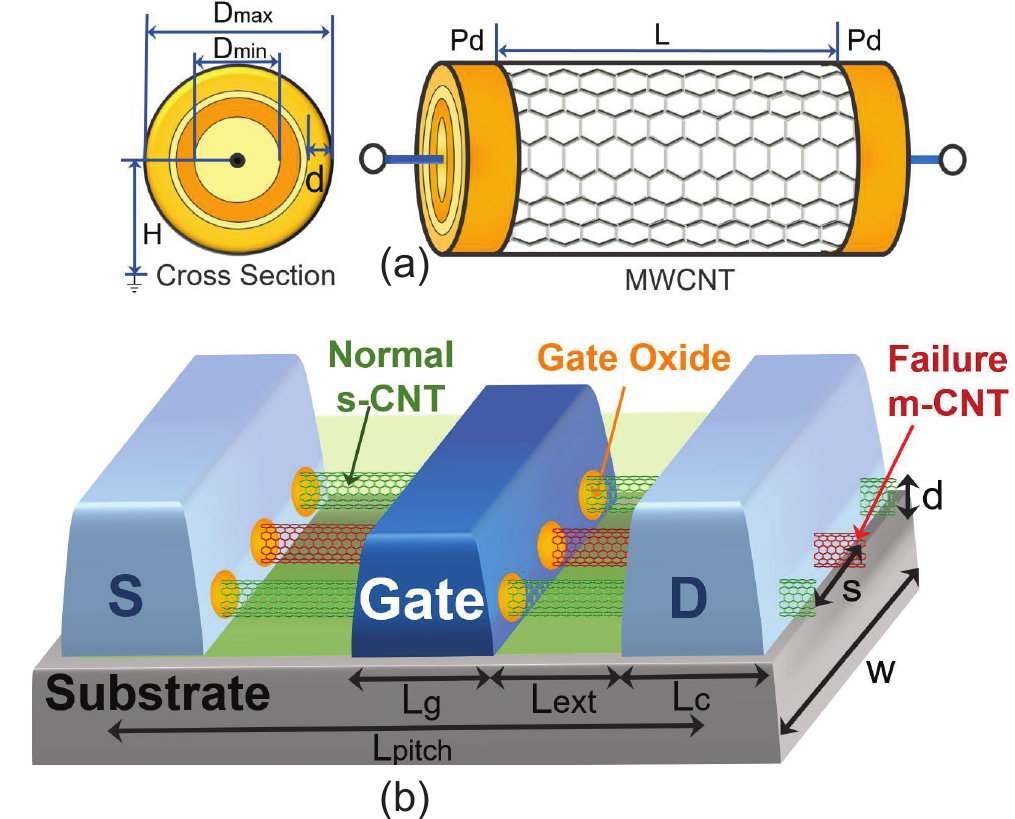}
	\caption{(a) Cross-sectional view and 3D view of an MWCNT structure (b) 3D view of a CNFET structure}
	\label{fig:back}
        \vspace{-0.6cm}
\end{figure}

Field-programmable gate arrays (FPGAs) have been the most popular reconfigurable fabrics in the past few decades and are widely used in numerous commercial applications \cite{b1, b1.1}. The high programmability of FPGAs is implemented by FPGAs’ interconnects and configurable logic. However, the resistivity of copper (Cu) interconnect increases due to electron surface scattering. Besides, these configurable logic blocks built with conventional MOSFETs suffer from large area overheads and high leakage power \cite{b2}. As shown in Fig. \ref{fig:back} (a), electromigration-resistant multi-walled carbon nanotube (MWCNTs) is a prospective candidates to replace Cu interconnect due to its superior conductivity and ampacity \cite{b3}. Moreover, as shown in Fig. \ref{fig:back} (b), carbon nanotube-based field-effect transistor (CNFET) is a promising alternative to MOSFET due to the extremely low power dissipation and high endurance \cite{b4}  \cite{b4.1}  \cite{b4.2}. It shows that CNT-based FPGAs can provide 2.67$\times$ performance improvement compared with CMOS-based FPGAs at the same technology node \cite{b5}. Recently, a microprocessor called N3XT is introduced, in which the logic and peripheral circuits are manufactured using CNT technology, which can achieve ~850$\times$ EDP (Energy-Delay product) improvement compared to MOSFET technology\cite{b0}.

While CNFET has many benefits compared with conventional MOSFETs, it can be widely adopted only if the design and manufacturing cost are commercially viable \cite{b6}. One of major challenges on the path to large-scale CNT-FPGAs is the process variation of multi-walled carbon nanotube (MWCNT) bundled interconnects \cite{b7}. Growing CNTs introduces several feature size variations (e.g., shell diameter and chirality), which affect the propagation delay of signals \cite{b3}. Many delay-fault testing methods for FPGAs are supported by built-in self-test (BIST) \cite{b7.01} structures, such as \cite{b7.1}~\cite{b7.2}. These works compared timing differences among the paths under test (PUTs). However, the test accuracy was degraded due to the uncertainty introduced by clock skew.

Another major challenge for CNT-based FPGAs is the presence of metallic CNTs (m-CNTs) in the channels of CNFETs devices \cite{b4}. As shown in Fig. \ref{fig:back} (b), the normal semiconducting CNTs (s-CNTs) are grown together with m-CNTs in the fabrication process. The s-CNTs are promising channel materials for building CNFETs, while an m-CNT can induce a stuck-on fault in a CNFET device, leading to the malfunction of circuits. Since an m-CNT can be as long as hundreds of micrometers \cite{b8}, a large number of correlated blocks can be faulty blocks once m-CNTs appear in CNT-based programmable gate arrays. The key open-ended problem of efficient testing is not solved satisfactorily in existing works \cite{b8.1}. Therefore, it is of great significance to propose fast and effective test techniques for CNT-based FPGAs. Moreover, it is necessary to propose novel redundant architectures to repair faulty components containing m-CNTs in order to improve yield further.

In this article, combining with the unique fault patterns in carbon nanotube-based programmable gate arrays, we propose a novel delay testing technique for multi-walled carbon nanotube (MWCNT) interconnects, and a fast diagnosis of the faulty segments in configurable logic blocks (CLBs). Besides, a redundant spare row sharing architecture is proposed to repair faulty segments. The main contributions of this work can be summarized as follows.

\begin{enumerate}
	\item[1.] We propose a ring oscillator (RO) based testing technique to identify delay faults on MWCNT interconnects.
	\hfill
	\item[2.] Then, we establish the fault models induced by metallic CNTs (m-CNTs) that may occur in a configurable logic block (CLB), and we propose the carry chain testing methodology supplementing the traditional test configurations. An improved testing circuit design based on a lookup table (LUT) in a CLB is also explored to speed up the delay fault testing.
	\hfill
	\item[3.] Considering these correlated faulty CLBs induced by m-CNTs, an effective technique to diagnose the faulty segments is proposed.
	\item[4.] Considering these faulty segments induced by m-CNTs, we propose a redundant architecture to repair the faulty segments effectively.
\end{enumerate}

The rest of this article is organized as follows. Section 2 introduces the background of the CNT-based FPGAs and the related work. Section 3  presents the MWCNT-based delay fault testing, illustrates the test technique for a single CLB considering m-CNT induced defects, and introduces the diagnosis methodology for the overall CLBs. Section 4 describes the redundant architecture of m-CNT-induced faulty tiles. Section 5 presents the experimental evaluations of our proposed testing and fault tolerant techniques. Section 6 concludes the paper.

\section{Preliminaries and Motivation}
\label{sec:why}
In this section, we introduce the preliminaries of CNT technology. Then, we present the motivation of the CNT-based FPGA architecture.
\subsection{Preliminaries}
\subsubsection{MWCNT and CNFET}
The multi-walled carbon nanotube (MWCNT) interconnect is considered as a promising alternative to Cu interconnect in terms of performance. MWCNT has many superior properties such as near ballistic transport, high conductivity and ampacity. As shown in Fig.~\ref{fig:back} (a), an MWCNT interconnect has several concentric shells with diameters ranging from several nanometers to tens of nanometers~\cite{b7}.

The structure and operation of a CNT-based field-effect transistor (CNFET) are analogous to those of a CMOS device. As shown in Fig.~\ref{fig:back} (b), semiconducting CNTs (s-CNT) form the conducting channel between source and drain, and can be controlled by a gate electrode. Based on intrinsic CV/I gate delay, CNFET devices can be up to 13× and 6× faster than pMOS and nMOS devices with the same gate length~\cite{b9.1}. 

In this work, we assume the MWCNTs serve as the interconnects, and CNFETs are used for transistors to construct the CNT-based FPGA.

\begin{figure}
	\centering
	\includegraphics[width=3.2in]{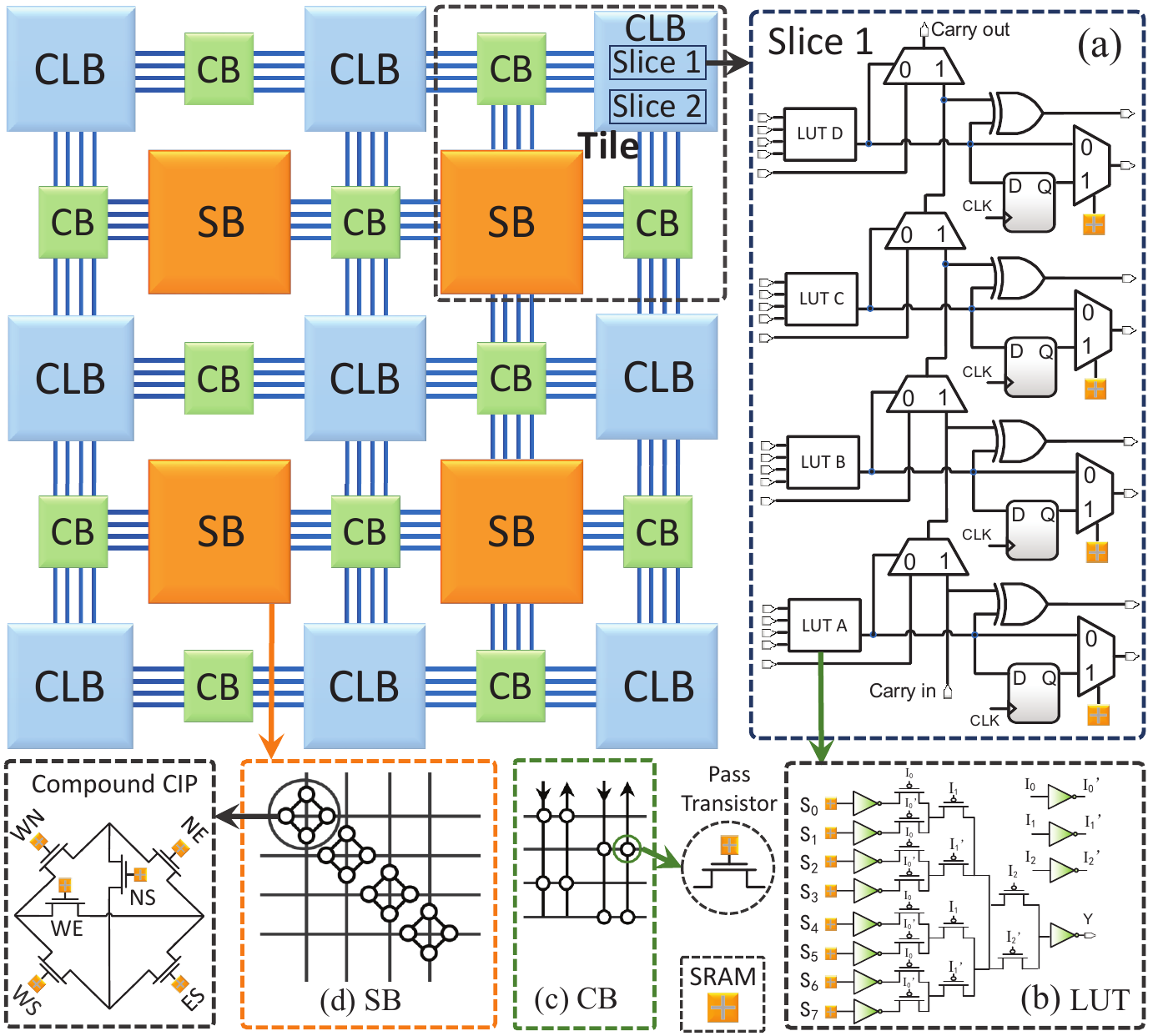}
	\caption{The CNT-based FPGA architecture  (a)The diagram of a slice in a CLB, which contains LUTs, carry chain and triggers  (b) The CNT-based LUT  (c) The connection block  (d) The switch block }
	\label{fig:fpga}
    \vspace{-0.5cm}
\end{figure}

\subsubsection{CNT-based FPGA Architecture}
Modern FPGAs typically use an island-based architecture, which is mainly composed of configurable logic blocks (CLBs) connected via programmable connection blocks (CBs) and switch blocks (SBs) \cite{b10}, and a tile is composed of a CLB connected via two CBs and a SB, as shown in Fig. \ref{fig:fpga}. A CLB consists of the combinational and sequential elements that are needed to implement specific logic functions. Static Random-Access Memory (SRAM) is used to store truth table in a lookup table (LUT) \cite{b10.01,b10.02} in a CLB. In a CBs/SBs, SRAM stores the configuration bits to control interconnects. The selection signals of multiplexers are also controlled by SRAM.

Fig. \ref{fig:fpga} (a) shows the schematic of a slice in a CLB, the programmability of a CLB is controlled by the lookup table (LUT). A slice is composed of LUTs, carry chain, flip-flops, and some multiplexers (MUXs) used to select the immediate output or registered output. An LUT with K inputs can implement any K-input Boolean functions. As shown in Fig. \ref{fig:fpga} (b), for a K-input LUT, $2^{K}$ SRAM cells are used to store the configuration bits, and a $2^{K}$-to-1 MUX is used to select the output bit.

In Fig. \ref{fig:fpga} (c), a CB consists of a few programmable switches and MWCNT interconnects. It serves to connect the interconnect channels and the inputs/outputs of CLBs. The pass transistors (PTs) instead of transmission gates are commonly used to implement programmable interconnects in FPGAs because each PT requires only one P-CNFET.

As shown in Fig. \ref{fig:fpga} (d), for each SB, its connection is controlled by six configurable interconnect points (CIPs). Individual CIP in a switch blocks is denoted by the relative directions of two to-be-connected wires. For example, a WN CIP connects a wire located in the west (W) to another wire in the north (N). To provide high flexible programmability, lots of programmable CLBs, SBs and CBs are integrated in an FPGA.

\subsection{Motivation}
The use of ring oscillators is an effective technique to measure variations in FPGA manufacturing processes. Li \textsl{et al.} used an array of ring oscillators in an FPGA to measure the gate length variation \cite{b7.3}, which was then used to improve the fabrication process and reduce the negative effect of process variations on circuit performance.
For the traditional testing scheme of CMOS-based CLBs, according to horizontal testing and vertical testing \cite{b10.1}, the faulty CLBs can be identified by intersecting faulty columns with faulty rows. But this scheme cannot test the cascaded faulty CLBs effectively. In the technique presented in \cite{b7.2}, every CLB used in the mapped design is reconfigured as transparent logic to construct scan chains. Also, fanout branches of a net are tested in different test configurations, resulting in a number of test scenarios. Due to the complexity of configuration generation algorithm, it cannot be applied to large designs.

Besides, although m-CNTs can be removed by electrical burning \cite{b12}, sorting \cite{b12.1}, and selective etching \cite{b12.2}, these techniques can not achieve perfect metallic CNT removal. The m-CNT removal process may also create an open circuit if some s-CNTs under the FET are removed (see Fig. \ref{fig:process} (b)). Therefore, testing unique faults on CNT-based FPGA becomes increasingly important to guarantee the performance and the yield.

In addition, for the CMOS-based faulty memory, it is common to use the Built-in Self-Repair (BISR) method \cite{b12.3}, which is a good way to repair faulty memory rows or columns and increase memory yield. But this method is suitable for discrete random failures, and not effective for continuous failures caused by random distribution of m-CNTs in CNFETs. Considering the CNT-based faulty SRAMs, the adjacent CLBs sharing technology (SSS-2) based on Divide BitLine (DBL) was proposed to repair faulty segments \cite{b12.4}. But when the number of CLBs is small, the repair rate is not satisfied. Furthermore, this method is only suitable for SRAM repair, and will cause significant overhead when applied for tile-based system level fault tolerance.

\section{Ring Oscillator-based Delay Fault Testing}
\label{sec:man}
In this section, we explore the delay fault of MWCNT interconnects and propose a Ring Oscillator (RO)-based BIST scheme, which can effectively test delay faults  induced by the MWCNT interconnects.

\begin{figure}[t]
	\centering
	\subfloat[]{
		\includegraphics[width=1.55in]{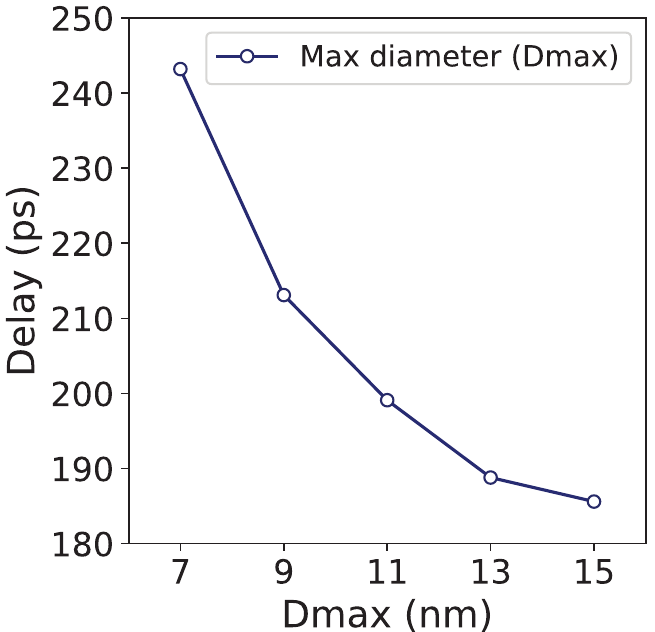}}
	\subfloat[]{
		\includegraphics[width=1.55in]{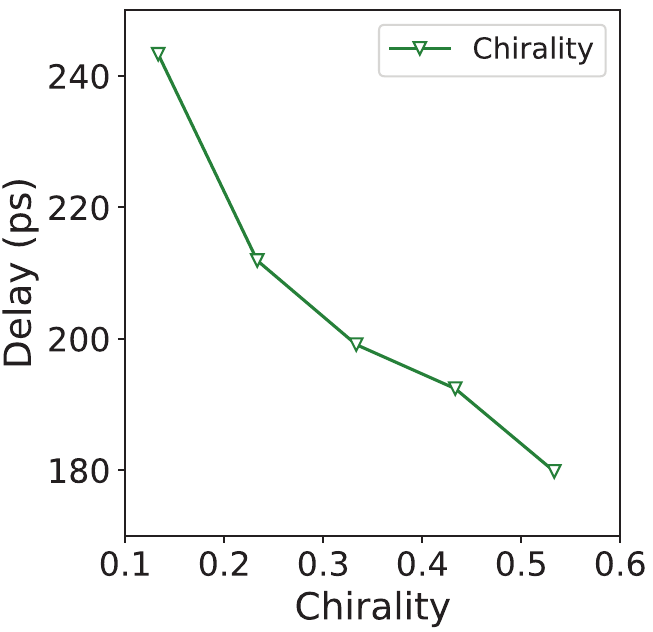}}
	\caption{The delay between two adjacent CNT-based CLBs with (a) Dmax variation (b) chirality variation}
	\label{fig:dmax}
    \vspace{-0.4cm}
\end{figure}
\subsection{The Delay Fault of MWCNT Interconnects}
As the line width and interconnect pitch scale down to nanometers, the propagation delay becomes a major performance concern. Related work shows that the CNT diameter and chirality variations play essential roles in determining the performance of MWCNT interconnects \cite{b3}. 

To evaluate the effects of above parameters on MWCNT delay, we plot the delay between two adjacent CNT-based CLBs with MWCNT variation settings as in \cite{b3}. As shown in Fig. \ref{fig:dmax} (a), delay between two adjacent CLBs with the $D_{max}$ variation decreases significantly as the diameter increases. Similarly, as the chirality of MWCNT is improved (see Fig. \ref{fig:dmax} (b)), the delay between two adjacent CLBs is also reduced. Nearly 37\% of delay improvement is observed when chirality is changed from 0.33 (without any chirality optimization during fabrication) to 0.53 (53\% metallic CNT).

We evaluate the timing variations due to diameter ($D_{max}$) and chirality variations by Monte Carlo (MC) simulations. We assume the $D_{max}$ variation obeys Gaussian distribution $N$~(11$nm$, $1.65^{2}nm^{2}$), and the chirality of each shell obey Bernoulli distribution with each shell of 1/3 probability to be metallic \cite{b3}.

\begin{figure}
	\centering
	\includegraphics[width=0.8\linewidth]{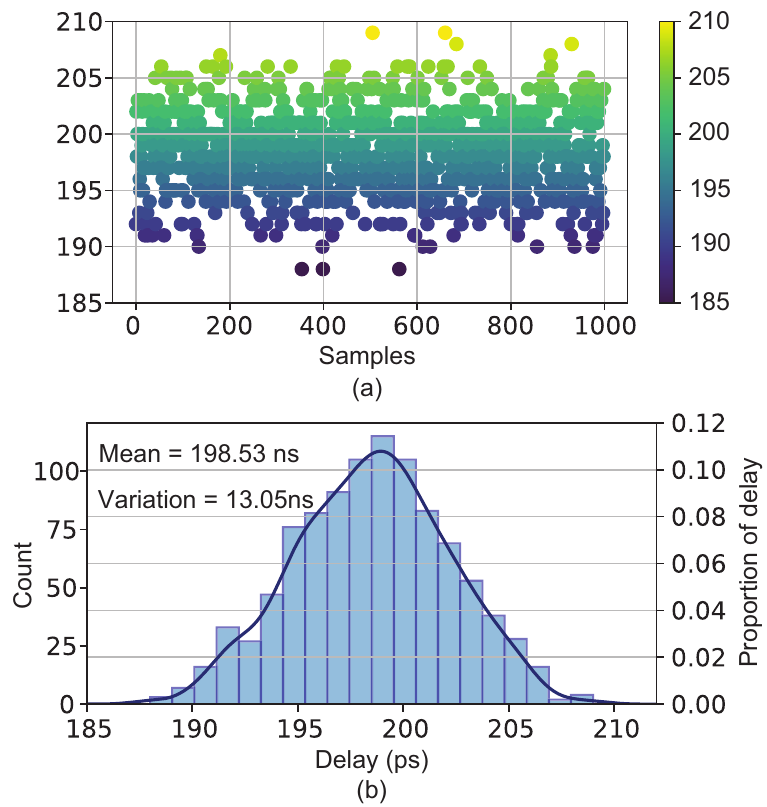}
	\vspace{-0.1cm}
	\caption{~The Monte Carlo simulation of propagation delay between two adjacent CLBs considering MWCNT process variations.}
	\label{fig:monte}
	\vspace{-0.6cm}
\end{figure}

The interconnect delay variation between two adjacent CLBs is shown in Fig. \ref{fig:monte}. We can observe that a few interconnect paths exhibit large delay variations. As the FPGA fabrication technology migrates to deep sub-micron regime, the impact of delay faults on interconnect paths will become more acute \cite{b7.1}.

\subsection{RO-based Delay Fault Testing of MWCNT Interconnects}
Routing resources consist of wire segments that are connected or disconnected by configurable interconnect points (CIPs). In this work, the ROs are constructed to measure the delay of MWCNT interconnect path. Note that XOR tree-based testing structures were used to detect delay faults of ASICs in prior works \cite{b17}, but this method did not deal with the application to FPGAs.

\begin{figure}[!h]
	\centering
	\includegraphics[width=1\linewidth]{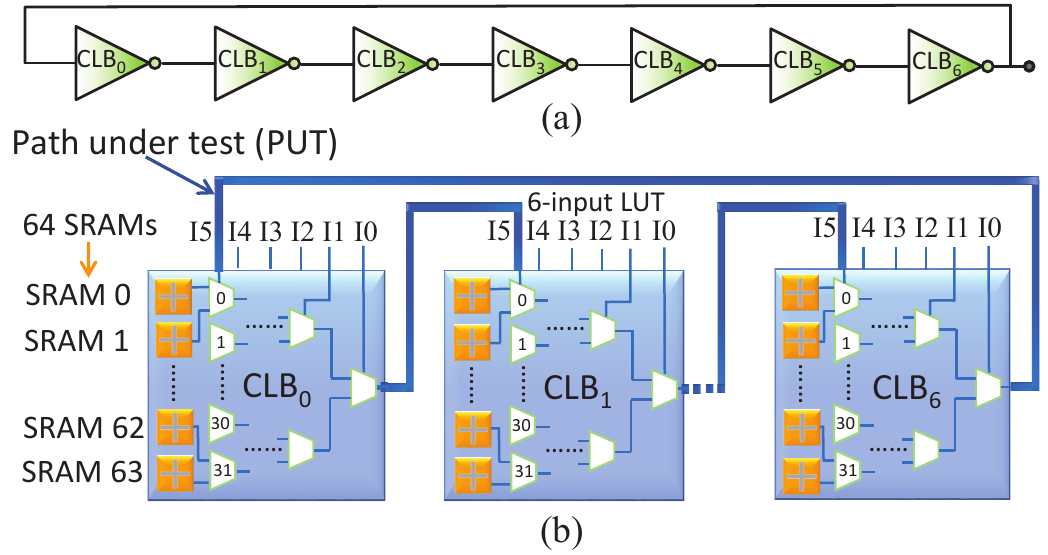}
	\vspace{-0.5cm}
	\caption{(a) The ring oscillator structure with 7-stages (b) Ring oscillator constructed by seven 6-input LUTs}
	\label{fig:ro}
	\vspace{-0.2cm}
\end{figure}

A RO is a circuit that consists of an odd number of inverting logic stages connected in series to form a closed-loop chain. An example where each stage consists of an inverter is shown in Fig. \ref{fig:ro}(a) The oscillating period is twice the sum of the propagation delay of all elements that compose the loop. ROs can be mapped on FPGAs using LUTs to measure the propagation delay of MWCNT interconnects. 

Fig. \ref{fig:ro}(b) illustrates a possible formation of a ring oscillator with 7-stages using 6-input LUTs. A 6-input LUT consists of 64 SRAMs and a 64:1 multiplexer. A tree of 2:1 multiplexers has been used to build the 64:1 multiplexer. Any 6-input Boolean function can be realized by setting the truth table values in the SRAMs, where the output is determined by the logic values of three-level hierarchical selectors (I5, I4, I3, I2, I1, and I0). The output of each LUT is connected to selection bits of the next LUT to form a closed-loop chain. For each test configuration, we denote the frequency of each 7-stage RO as $f_{1}$, $f_{2}$, $f_{3}$, ......., $f_{N^{2}/7}$ respectively. $N^{2}$ represents the number of tiles in the FPGA array. Proper mapping of SRAM cell data and MUX selection bits are required to obtain the oscillation behavior from LUTs.

A Boolean function that implements inverter logic is needed. Standard logic XNOR and XOR can be used as an inverter when one of the inputs is considered as RO input while other inputs remain stable. An example of XNOR-based mapping for 6-input LUT is illustrated in Table \ref{tab:rot}, the operator ‘$\odot$’ represents the XNOR operation. In Section \ref{subsec:re}, we map odd LUTs (7 LUTs) into a ring oscillator as a RO cell. LUT input pin I0 serves as input to inverter logic, and XNOR-based ROs can satisfy the proposed testing requirement.

Compared with traditional scan-based or MUX-based delay test techniques, the proposed RO-based structure eliminates the need for external pattern generation and I/O measurement, and achieves higher fault coverage with significantly lower hardware overhead.

While prior techniques such as scan/launch methods require complex scan chains and dedicated timing paths, our approach leverages on-chip reconfigurable LUTs to form oscillators that directly measure interconnect delay as a frequency shift.

\begin{table} [!h]
	\vspace{-0.2cm}
	\centering
	\caption{Formation of oscillating path with 6-input LUTs using XNOR logic function}
	\begin{tabular}{c}
		\includegraphics[width=0.9\linewidth]{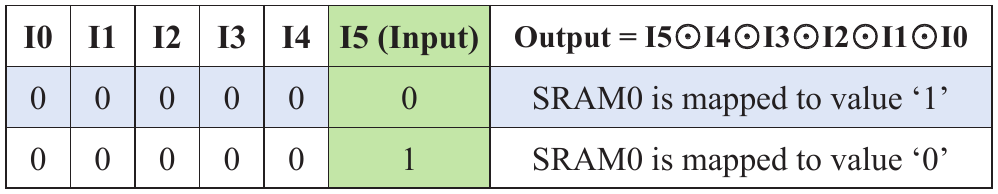}
	\end{tabular}
	\label{tab:rot}
	\vspace{-0.4cm}
\end{table}

\section{Fault Testing Methodologies for Configurable Logic Blocks}
\label{sec:part}
In this section, we first analyze fault models induced by m-CNTs that may occur in a CLB. Then, we propose the test technique for the carry chain in a CLB, and a test circuit design based on an LUT is also explored to speed up the fault testing. Finally, we propose a technique to diagnose and repair the faulty segments within an FPGA tile.

\subsection{M-CNT Induced Fault Model}
\label{subsec:re}

\begin{figure}[!t]
	\centering
	\includegraphics[width=1\linewidth]{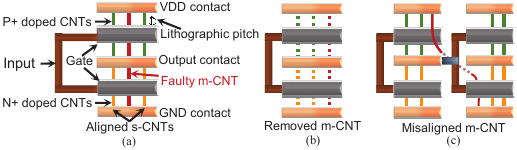}
	\vspace{-0.4cm}
	\caption{m-CNT in CNT-based inverter (a) m-CNT leading to a short fault (b) s-CNT removal leading to an open fault (c) Misaligned CNTs in a two-stage inverter}
	\label{fig:process}
	\vspace{-0.4cm}
\end{figure}

As mentioned in Section \ref{sec:man}, s-CNTs form the channel of CNFET. However, m-CNTs may grow together with normal s-CNT, and a typical CNT synthesis process yields roughly 3\%-33\% of m-CNTs. As shown in Fig. \ref{fig:process} (a), CNFET containing such m-CNTs is no longer controlled by the gate, which can lead to a short defect between source and drain, and cause a failure in the CNFET. Moreover, m-CNTs affect the current flowing through a CNFET when it is switched on ($I_{on}$) and off ($I_{off}$). If the ratio $\frac{I_{on}}{I_{off}}$ of a CNFET cannot reach a designated threshold value, the CNFET is regarded as having an open fault. 

Considering a representative design scenario, an FPGA contains hundreds of thousands of tiles, with a 5\% m-CNT probability (a typical CNT synthesis process results in a percentage of m-CNTs in the range 3\%-33\%) and 99.99\% m-CNT removal percentage \cite{b12}. There may be dozens of faulty tiles in the CNT-based FPGA. The growing length of a CNT can be as long as hundreds of micron meters, an m-CNT can lead to a few correlated faulty CLBs; a misaligned m-CNT may span different rows, which can result in faulty segments in CNT-based FPGAs, and cause several cascaded CLBs to malfunction. The m-CNT can also have length and angle variations, and the induced faults are highly correlated with the CNT growing direction.

As mentioned above, for the CNT-based logic circuits, the growth of m-CNTs may cause incorrect circuit functionality. A CLB in the CNT-based programmable gate array is mainly composed of LUTs (including SRAMs, multiplexers, etc.), carry chains and triggers. An LUT is typically built out of SRAM bits that hold mapped values and a set of multiplexers (MUX) to select the bit that drives the LUT output. Next we analyze the fault model of an LUT, and we verify the fault models of different LUT components according to \cite{b15}. 

\begin{figure}[t!]
	\centering
	\includegraphics[width=1\linewidth]{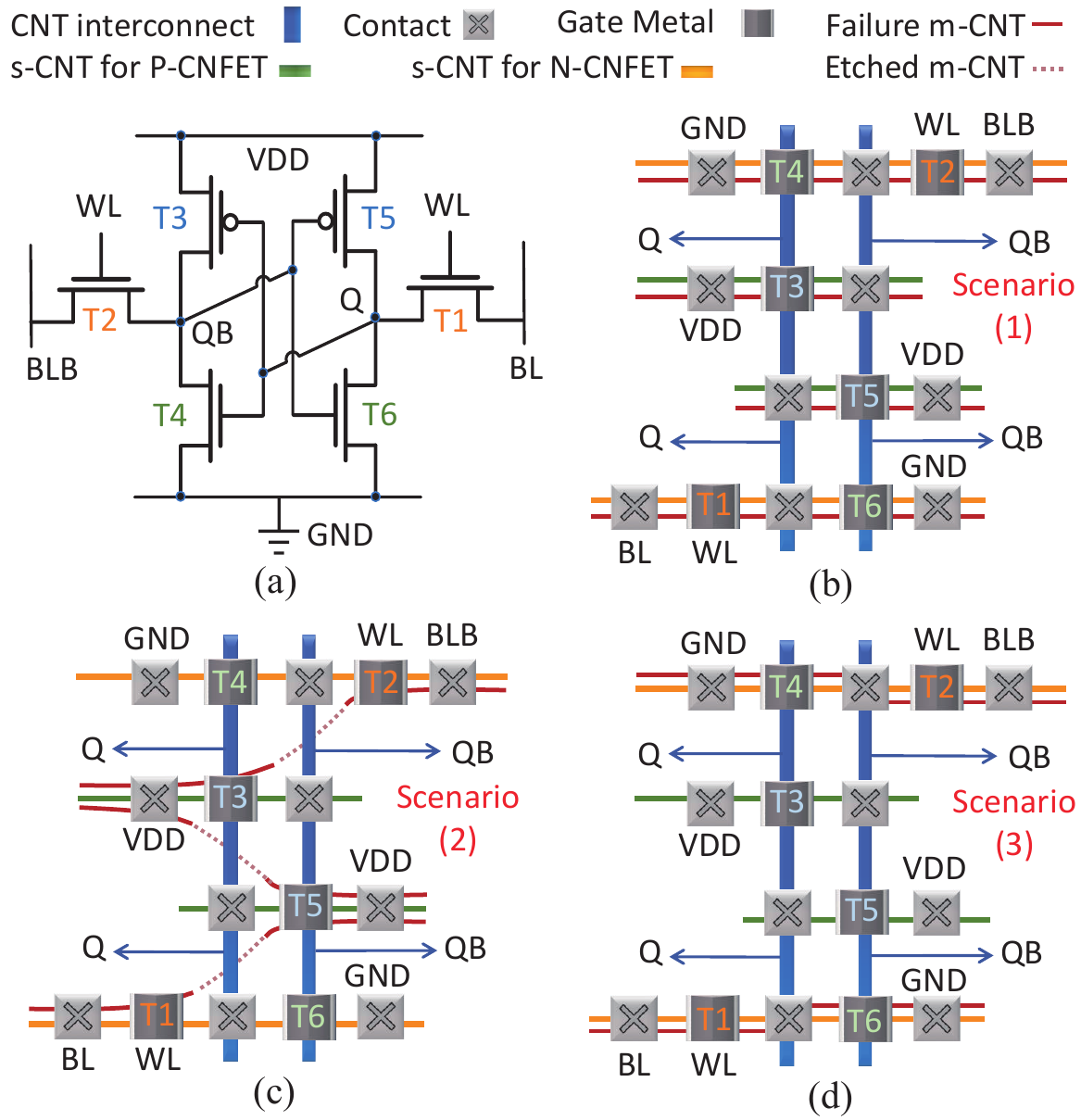}
	\vspace{-0.3cm}
	\caption{~(a) The circuit schematic of a CNT-based SRAM cell. (b) The faulty layout of a SRAM cell induced by m-CNTs, which grow together with s-CNTs. (c) The faulty layout of a SRAM cell induced by misaligned m-CNTs. (d) The faulty layout of a SRAM due to the length variation of m-CNTs.}
	\label{fig:sramt}
	\vspace{-0.6cm}
\end{figure}

\textsl{1) SRAM fault model:} As shown in Fig. \ref{fig:sramt} (b), we consider the following scenarios: In scenario (1), an m-CNT passes through two horizontal CNFETs in a row, e.g., $T_{2}$. In scenario (2), a misaligned m-CNT affects non-vertical CNFETs in two CNT bundles, e.g., $T_{2}$ and $T_{3}$. In scenario (3), an m-CNT terminates after it passes through one CNFET, e.g., $T_{1}$. Note that the etched m-CNT is drawn with the light red dotted line.

A CNT-based SRAM cell may exhibit various types of faults, depending on the locations of m-CNTs in the SRAM. Table \ref{tab:sramb} summarizes the fault models considering typical m-CNT induced faults at the transistor level. The first column describes different scenarios mentioned above. In the second column, ‘X–Y’ means two shorted positions, ‘X’ and ‘Y’. The third column refers to the transistor label in the SRAM that suffers from the short defect, and the last column lists faults to be detected. We observe that all faults can be modeled as the conventional stuck-at fault.

\begin{table} [!t]
	\centering
	\caption{The fault models of CNT-based SRAM.}
	\includegraphics[width=1\linewidth]{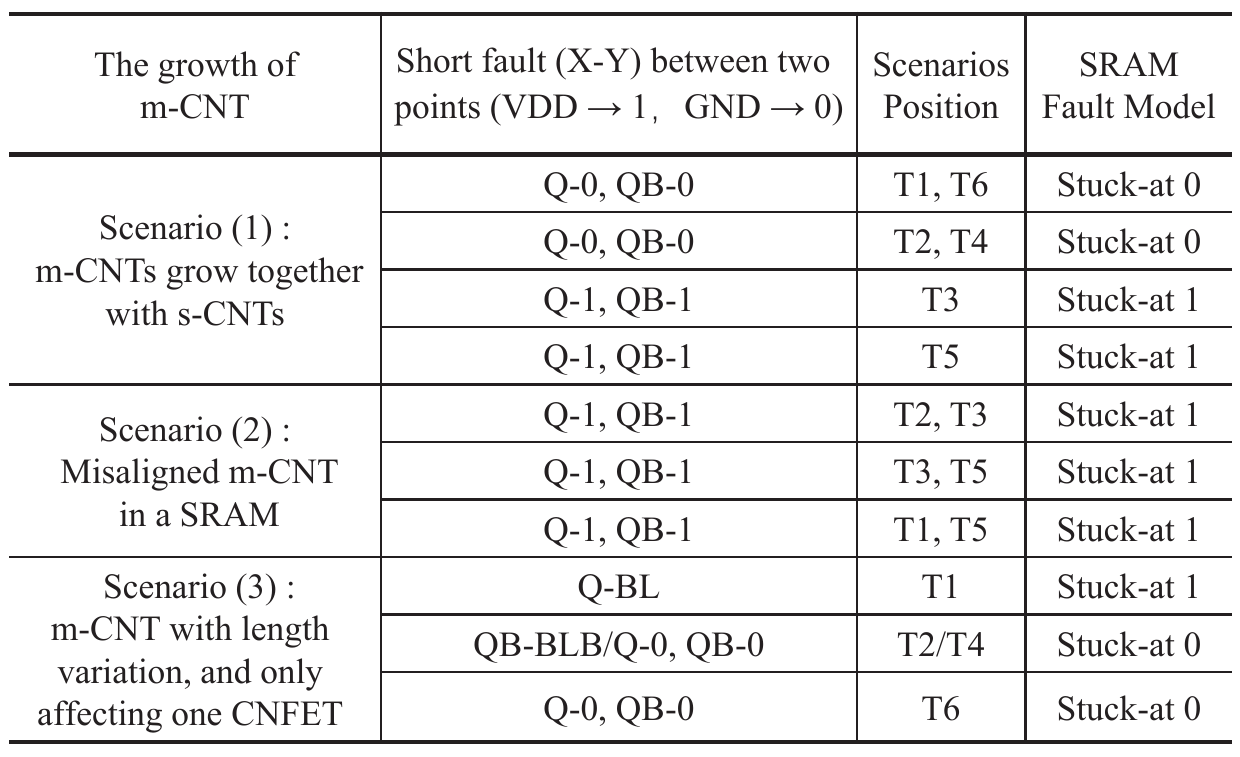}
	\label{tab:sramb}
	\vspace{-0.7cm}
\end{table}

\begin{figure}[!t]
	\centering
	\includegraphics[width=1\linewidth]{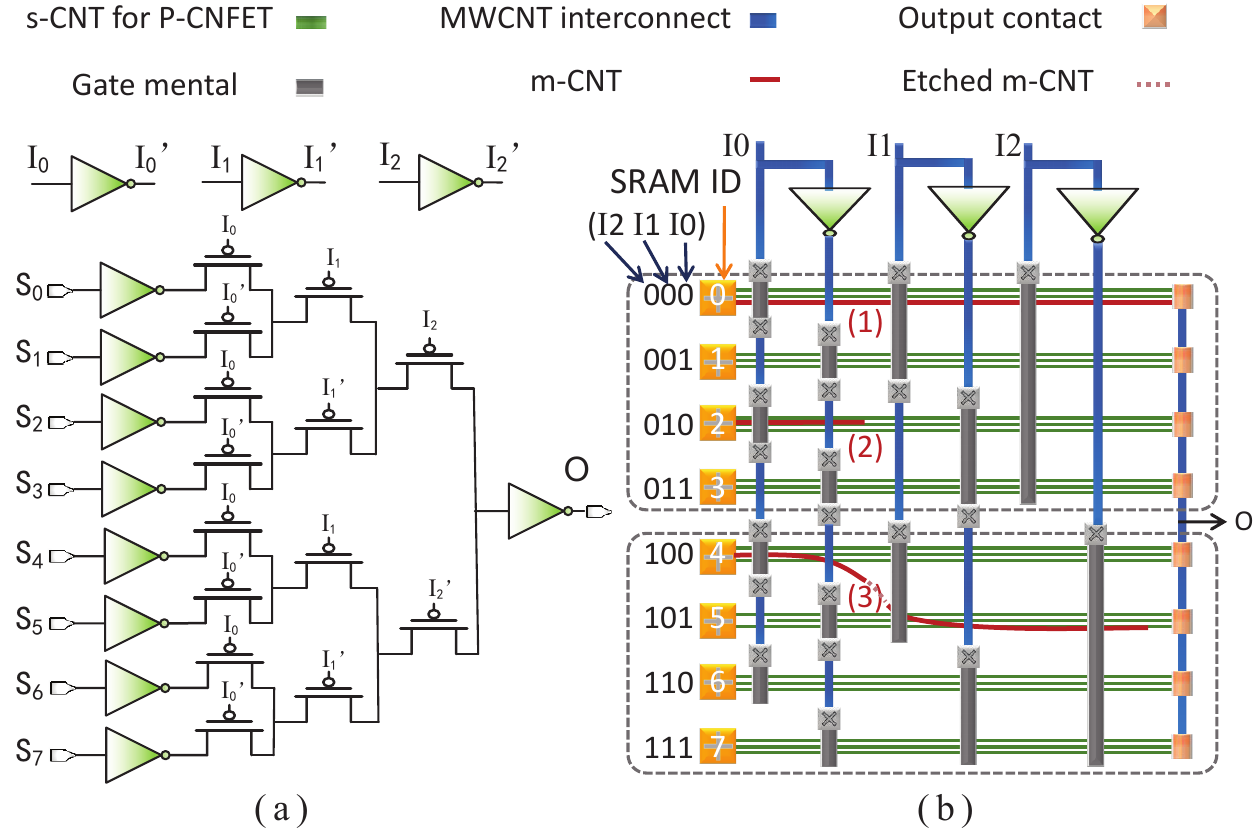}
	\vspace{-0.4cm}
	\caption{(a) The circuit schematic of a MUX. (b) The faulty layout of a MUX induced by m-CNTs.}
	\label{fig:muxt}
	\vspace{-0.6cm}
\end{figure}

\textsl{2) MUX fault model:} As shown in Fig. \ref{fig:muxt} (a), a 3-input LUT is composed of 8 SRAM cells and an 8:1 multiplexer. Any logic function of 3-inputs can be realized by setting the appropriate value in the SRAM cells and 3 level hierarchical selectors (I0, I1, I2). The CNT-based MUX can be built by P-CNFETs as transmission transistors. A 3-input LUT needs 8 CNT bundles as transmission paths. When scaled to K inputs, the LUT will contain $2^{K}$ CNT bundles.

The CNT-based MUX may also present various faults depending on positions of m-CNTs, and we analyze the following typical scenarios in Fig. \ref{fig:muxt} (b): In scenario (1), an m-CNT passes through a whole row, and affects all selection (S) signals, e.g., short between SRAM-0 and output, which leads the MUX to always output the value stored in SRAM-0. In scenario (2), an m-CNT terminates after it passes through one S signal, e.g., a short between SRAM-2 and a CNFET. When the SRAM-3 output is selected, it causes a wired-AND/OR fault of the values stored in SRAM-2 and SRAM-3. In scenario (3), a misaligned m-CNT affects transmission gates in two CNT bundles: the front m-CNT causes a wired-AND/OR fault of SRAM-4 and SRAM-5 when the SRAM-5 output is selected, the end part of the m-CNT also causes a wired-AND/OR fault when the output of SRAM-1/SRAM-3/SRAM-7 is selected (i.e., affects the output of SRAMs corresponding to $\overline{I2}$).

\subsection{Overview of Fault Types}
As mentioned in Section \ref{sec:why}, faults may occur in the LUT (including SRAM memories, a MUX and inverters), the carry chain logic (including MUXs and XOR gates) and triggers.

The fault type of a CNT-based SRAM can be considered as the conventional stuck-at faults. If a fault occurs in an SRAM cell, wrong SRAM value will be output. 

A multiplexer is a group of switches and only one switch is allowed to be on. The fault type of a CNT-based MUX can be regarded as short fault or the wired-AND (wired-OR) fault as mentioned in Section \ref{sec:man}.

For a trigger (D flip-flop), a fault may cause the trigger to receive wrong data or to be incapable of being set or reset.

\section{Problem Formulation and Analysis}
\label{5_1}
\subsection{Fault Testing for a Single CNT-based CLB}
In this work, we adopt the test technique for a single CLB proposed in \cite{b18} and we add the carry chain test in the CLB, which is not solved in \cite{b18}. Then, an improved design based on LUT is also proposed to speed up the fault testing.

\textsl{1) Universal test procedure:} The configuration SRAM memory cells (CMCs) are used to configure its logic functions. When programming an FPGA, we can load the bit patterns into CMCs \cite{b15.1}. Such a programming process is called a configuration. We denote the procedure for testing CLBs in CNT-FPGAs as the test session $TS_{CLB}$. Then, we represent $TS_{CLB}$ consisting of a configuration and input test patterns \cite{b7.4, b7.5} applied to the configuration as follows:
\begin{equation}
	TS_{CLB}=[(TC_{1}, Seq_{1}), (TC_{2}, Seq_{2}), ..., (TC_{k+1}, Seq_{k+1})]
	\label{eq:con}
\end{equation}
where $TC_{i}$ is the $i$th configuration, $TP$ is the bit patterns applied to $TC_{i}$, and k is the number of inputs of a LUT. For each test configuration, the number of TP is $2^{k}$, so the the minimal length of complete input sequence $Seq_{k+1}$ (the number of TPs per TC) applied for $TC_{i}$ can be expressed as
\begin{equation}
	\left|Seq_{1}\right|=\left|Seq_{2}\right|=\cdots=\left|Seq_{k+1}\right|=2^{k}
	\label{eq:con}
\end{equation}

Since an LUT can realize $2^{n}$ ($n=2^{k}$) different functions, it is impractical to test each function exhaustively~\cite{b18}.

\begin{figure}[!t]
	\centering
	\includegraphics[width=1\linewidth]{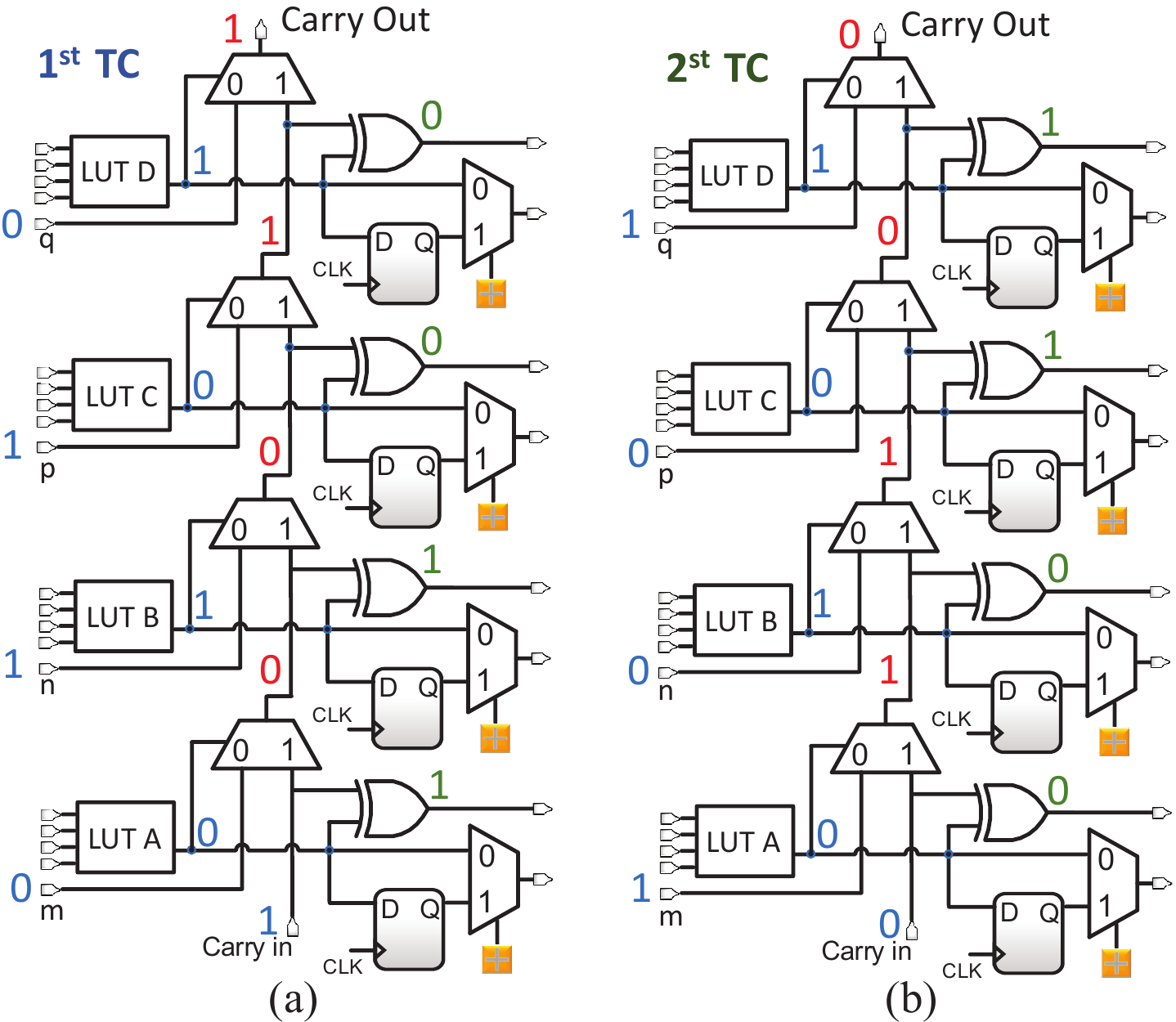}
	\vspace{-0.4cm}
	\caption{~The test configurations applied to the carry chain for fault detection. (a) The first test configuration. (b) The second test configuration.}
	\label{fig:cct}
    	\vspace{-0.4cm}
\end{figure}

\textsl{2) The proposed carry chain test:} With the modern FPGA technology, a dedicated carry chain is embedded in every CLB. The carry chain comprises basically MUXs and XOR gates to compute both the carry-out and the sum bits, respectively. Fig. \ref{fig:cct} shows a carry chain circuit and the associated LUTs \cite{b14}. 
The labels in Fig. \ref{fig:cct} indicate how to configure the outputs of LUTs and the external terminals. The detailed testing configuration is shown in Table \ref{tab:ccb}. At the first logic stage (i.e., the bottommost one), the MUX inputs are set to constant values `0' and `1', forcing it to propagate the output (`0') of LUT A to the next stage. So the output of the first MUX and XOR gate are constant values `0' and `1', respectively. Through the internal routing of the slice, the output of the first MUX is directly connected to the input of the second one. 

\begin{table} [!t]
	\centering
	\caption{~Two test configurations for the carry chain}
	\vspace{-0.1cm}
	\label{tab:ccb}
	\begin{tabular}{c}
		\includegraphics[width=0.85\linewidth]{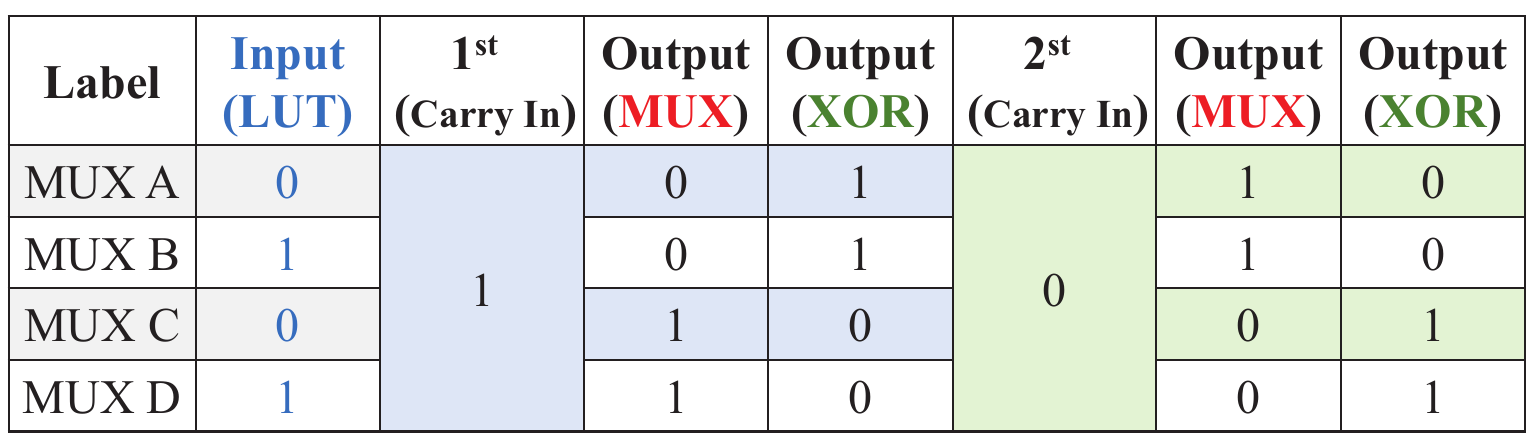}
	\end{tabular}
	\vspace{-0.6cm}
\end{table}

Then, the two inputs of the second stage’s MUX are now set to `1' and `0', respectively. As the selection signal (`1') is the output of LUT B, the second MUX is equivalent to calculating the XNOR function of the external signal (`1') and the signal (`0') from the first MUX. Note that at this moment MUXs in the carry chain are configured to perform XNOR operations. 
Hence, we can perform the computation of the XOR and XNOR functions of the entire carry chain. All stuck-at faults within the carry chain can be detected by configuring the MUX and XOR accordingly.

\begin{figure}[!t]
	\centering
	\includegraphics[width=1\linewidth]{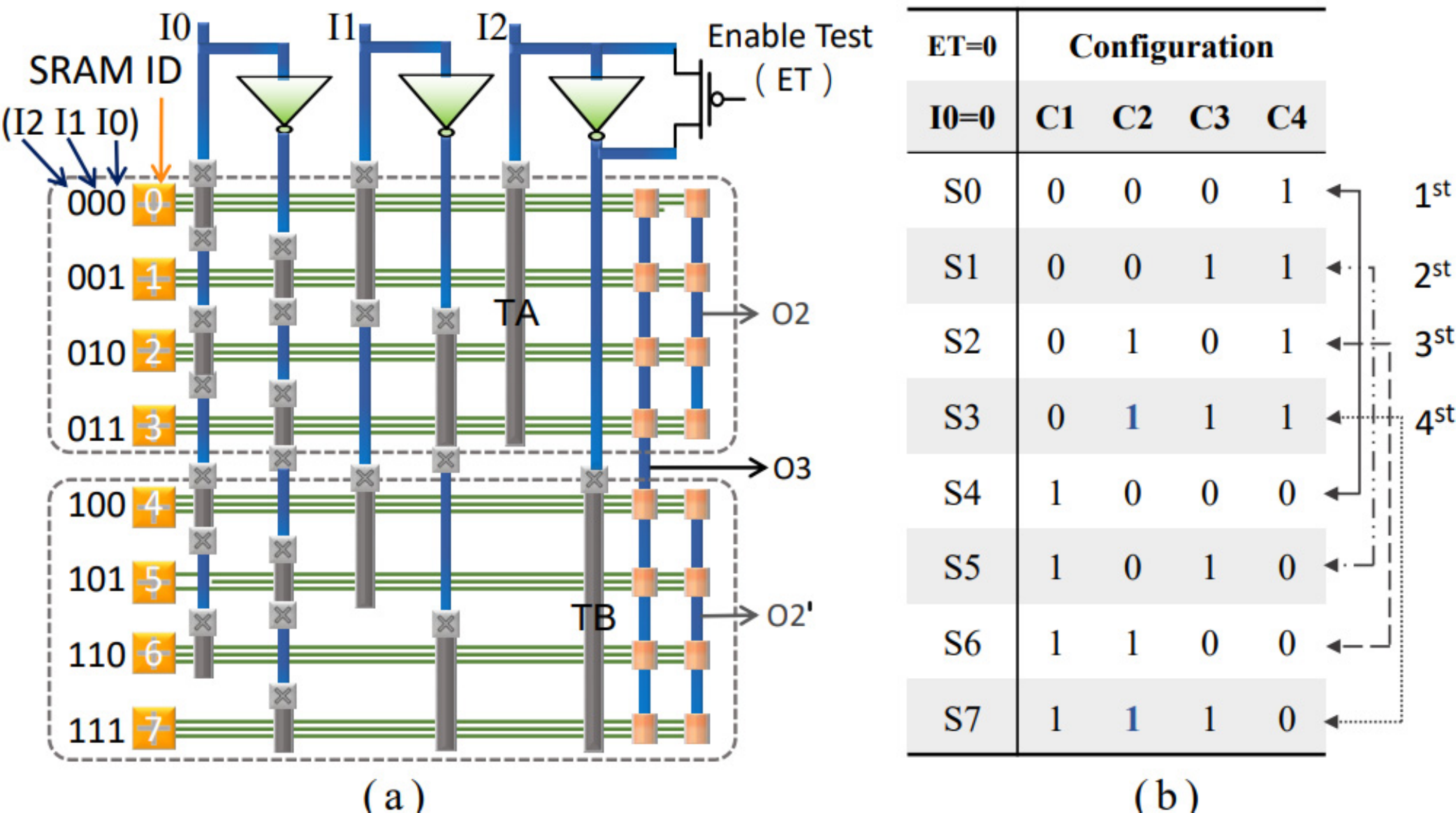}
	\vspace{-0.3cm}
	\caption{~(a) The proposed improved design of LUT. (b) The traditional test configuration scheme~\cite{b14}.}
	\label{fig:improve}
	\vspace{-0.6cm}
\end{figure}

\textsl{3) The proposed improved design for a LUT:} To speed up the testing application time for each configuration, an improved design based on an LUT in a CLB is proposed. As shown in Fig. \ref{fig:improve} (a), a P-CNFET is placed on the right of the inverter connected with the signal I2. The inverter enters the normal mode when ET=1, and enters the test mode when ET=0 (The inverter is shorted by the placed P-CNFET). 

In addition, contacts are placed after the pull-up and pull-down networks of the MUX, respectively. The MUX is divided into two networks, which can be tested in parallel, i.e., NW1 and NW2. For example, as shown in Fig. \ref{fig:improve} (b), the paths corresponding to SRAM-0 and SRAM-3 can be tested simultaneously in the $1^{st}$ group of test patterns, where the traditional eight test patterns are compressed into four patterns. However, when ET=0, I0=0, the pass transistors (TA and TB) corresponding to selected signal I0 are always on. Therefore, this technique cannot detect the stuck-on faults of these two gates. It needs to add a test pattern in configuration C2, i.e., (I2, I1, I0)=111. Because the values stored in SRAM-3 and SRAM-7 are all logic ‘1’, if any O2 or O2' outputs a logic value ‘1’ in this pattern, the transistor (TA/TB) is considered to have a stuck-on fault, and vice versa.

In summary, for one test session, the traditional method of testing each LUT requires k+1 test configurations with $2^{k}$ test patterns per configuration where k and $2^{k}$ represent the number of inputs and the configuration bits of a LUT, respectively. For example, as shown in Fig. \ref{fig:improve} (b), a 3-input LUT with 8 configuration bits requires 4 configurations and 8 test patterns per configuration. But for the proposed improved design of LUT, each LUT requires only 2 configurations and 4 test patterns per configuration in this case, so the test time is shorter than that of the traditional method.

\section{The Proposed Heuristic Algorithm}
\label{sec:five}

\begin{figure}[t]
	\centering
	\includegraphics[width=1\linewidth]{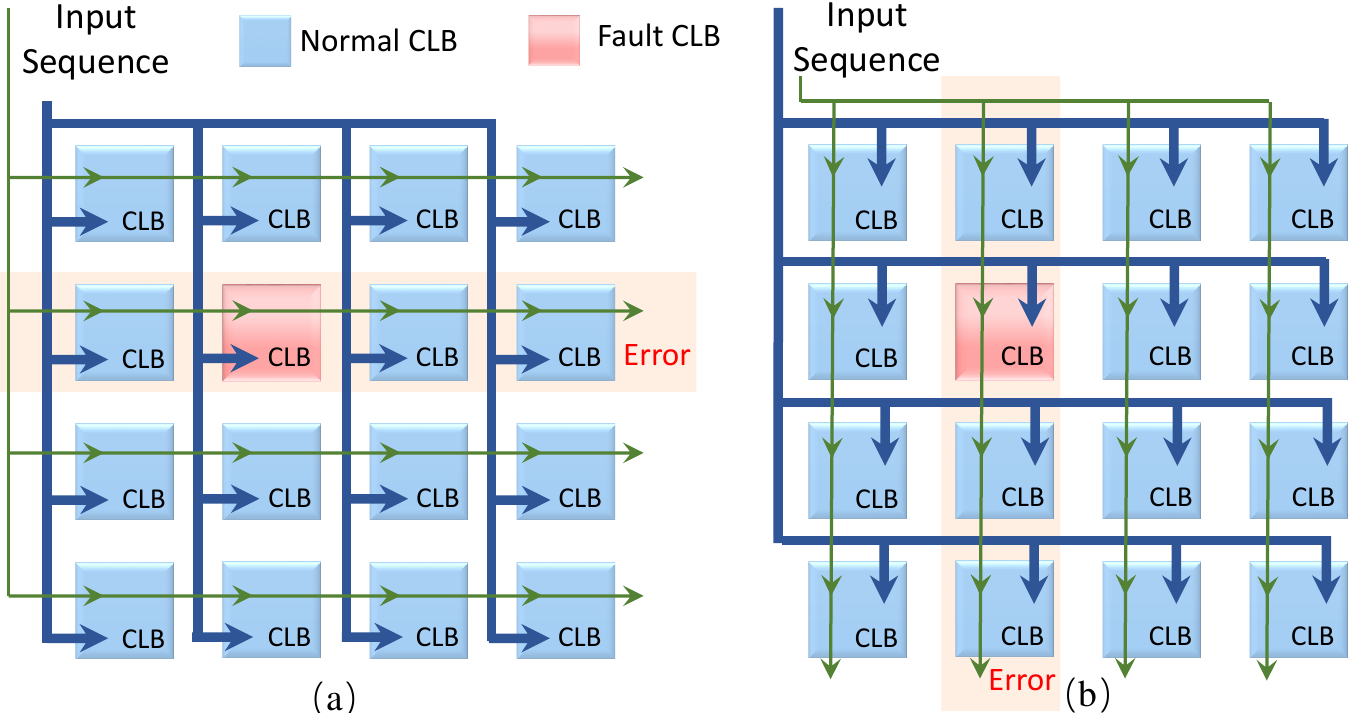}
	\vspace{-0.3cm}
	\caption{~The traditional method to diagnose the faulty CLBs~\cite{b27}.}
	\label{fig:trad}
	\vspace{-0.1cm}
\end{figure}

\textbf{Step I: Fault Testing Technique for the Overall CLB Array.}
As shown in Fig. \ref{fig:trad}, the traditional fault testing of CLBs contains two sessions: 1) horizontal test, the outputs of the rightmost CLBs are compared with correct responses to identify faulty rows. 2) vertical test, the outputs of the bottom CLBs are compared with correct responses to identify faulty columns. By intersecting faulty columns with faulty rows, the faulty cells can be identified \cite{b27}. 

However, the correlated fault CLBs induced by several hundred micrometers long m-CNT may span across several rows, resulting in correlated faults faults in these rows. The traditional test technique is only suitable for MOSFET-based FPGAs, and cannot test the cascaded faulty CLBs effectively. Note that diagnosing and locating the faulty CLBs of each row should be sensitized from the external IO ports \cite{b14}.

In this work, we explore the unique property of these cascaded faults induced by m-CNTs, and propose to make the test “jump” over CLBs. This can reduce the test overhead effectively. 

The idea of the recursive jump testing is to dynamically configure the direction and step size of each jump based on previous testing results until we locate both ends of the faulty segments. The algorithm is divided into two phases.

\begin{figure}[!t]
	\vspace{-0.2cm}
	\centering
	\includegraphics[width=0.99\linewidth]{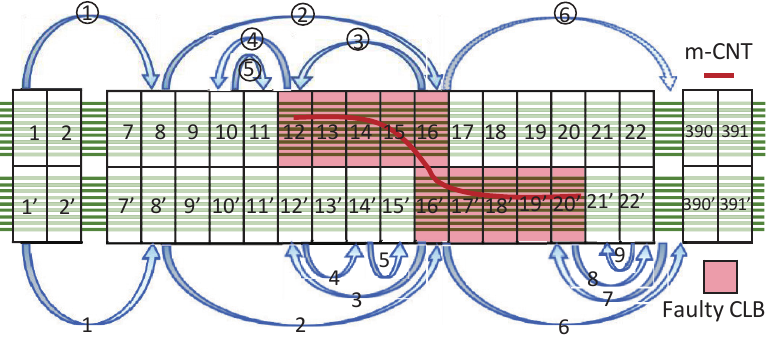}
	\vspace{-0.2cm}
	\caption{~An example illustrating recursive jump test.}
	\label{fig:rjt}
	\vspace{-0.2cm}
\end{figure}

\textsl{1) Initial phase:} In this phase, we jump with an initial jump step size through the columns along the rows, and record each test response of the current test (see the first and second jump in Fig. \ref{fig:rjt}). This phase aims to detect the faulty segment. Note that the jump step is always the same in the initial phase. Once the test response of the current CLB is different from the previous one, indicating that a faulty segment has been detected. Then, we enter the next phase denoted as the recursive phase (after the second jump).

\textsl{2) Recursive phase:} Note that the jump step in this phase is halved in each iteration. Once the test result of the jump is different from the previous one, the test jumps in the opposite direction, e.g., the third test jumps backward in Fig. \ref{fig:rjt}, but with only a half step size. If the test result of the third test is the same as the previous one, the test will jump forward again with the half jump step (see the fourth test). In our case, the test response of the fourth test is different from the third one, then we jump in the opposite direction and halve jump step size again (see the fifth test).

\begin{algorithm}[!t]  
	\caption{Recursive Test}  
	\label{alg:rja}  
	\begin{algorithmic}[1] 
		
		\Require Jump direction $Dir$, jump length $Step$, direction flag $Key$, row $i$ column $j$ CLB $C_{i,j}$,	recursive function Recursive($Dir$, $Step$, $Key$, $C_{i,j}$)
		\If {$Step$ = 1}
		\State Return;
		\Else
		\State $Step$=$Step/2$;
		\State $Key$ = ($C_{i,j}$ xor $C_{i,j+Dir \times Step}$) and Key
		\If {$Key$ = 1} 
		\State $Dir$ = -$Dir$;
		\State Recursive($Dir$,$Step$,$Key$,$C_{i,j+Dir \times Step}$);
		\Else
		\State Recursive($Dir$,$Step$,$Key$,$C_{i,j+Dir \times Step}$);
		\EndIf
		\EndIf
	\end{algorithmic}  
\end{algorithm}

This phase continues until we cannot divide the jump step further (e.g., the fifth test has jump step 1). The starting point of the faulty segment can be located at the moment. Consequently, we quit the recursive phase and continue the initial phase to detect the endpoint of the faulty segment. 

In summary, the pseudo-code of recursive testing algorithm is shown in Algorithm \ref{alg:rja}, where in the variable $Dir$ indicates the jump direction, $Step$ means the jump step size, $Key$ is to judge whether the two test responses before and after each recursion are equal, $C_{i,j}$ is the location of the CLB.

\textbf{Step II: Exploration of the Redundant Row Sharing Architecture.}
In this section, we firstly discuss the traditional method for repairing faulty segments in CNT-based FPGAs, and then explore the redundant architecture to repair faulty CNT-based FPGAs. 

Previous studies have proposed a rich set of methods to deal with faults in CNFET circuits \cite{b12.3}~\cite{b12.4}. What is currently missing is a way to increase the optimal amount of redundancy by adding redundant rows/columns and determining the appropriate sharing scheme to maximize yield while minimizing hardware overheads. According to the distribution of m-CNTs mentioned in Section \ref{sec:part}, the m-CNT grows in the direction of the column and may also span different rows.

\begin{figure}[!t]
	\centering
	\includegraphics[width=1\linewidth]{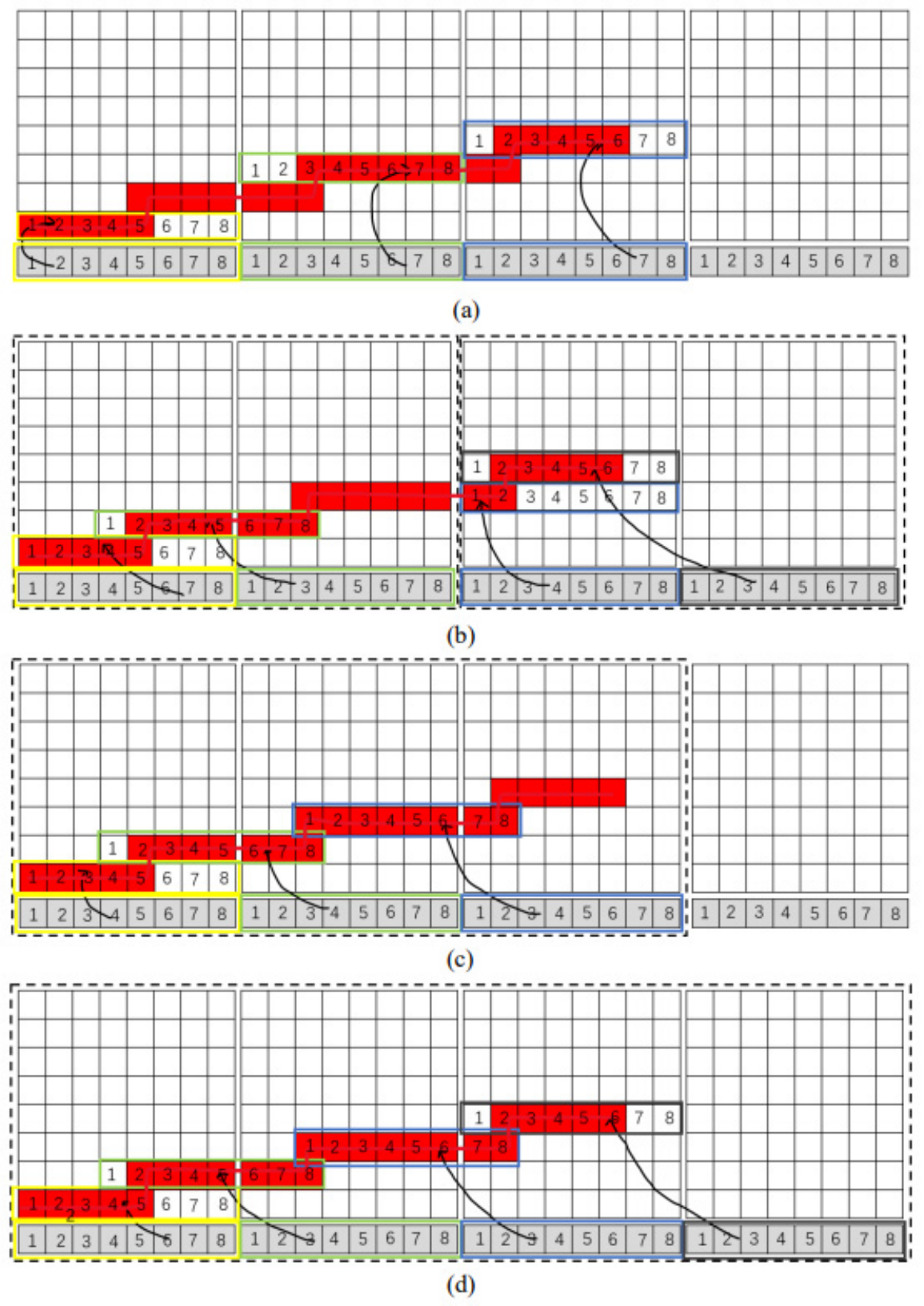}
	\vspace{-0.6cm}
	\caption{~(a) One 8×8 tile shares one spare row. (b) Two 8×8 tiles share two spare rows. (c) Three 8×8 tiles share three spare rows. (d) Four 8×8 tiles share four spare rows.}
	\label{fig:redundan}
	\vspace{-0.6cm}
\end{figure}

The traditional redundancy spare column/row repair scheme is only suitable for MOSFET-based FPGAs, and cannot effectively repair faulty tiles due to m-CNTs in the CNT-based FPGAs. In contrast to the discrete faults in CMOS-based RAM (which occur in a two-dimensional local area), the faulty segments caused by m-CNTs expand along the direction of CNT growth and may affect continuous column blocks. So in order to reduce fault tolerance overhead, we propose a redundant spare row sharing scheme to repair the faulty segments induced by m-CNTs. We divide total FPGA tiles into multiple small tiles and share alternative rows in adjacent tiles. The granularity of sharing is one spare row rather than with an entire tile, which can reduce hardware overhead significantly.

The basic concepts of the alternative row-sharing architecture are shown in Fig. \ref{fig:redundan}. Small tiles of multiple adjacent columns are grouped together to form a tile group that can share redundant spare rows. Within a tile group, any faulty segment in a small tile can be replaced by any spare row, and each spare row can span two small tiles for fault tolerance. The redundant spare row sharing schemes considered in our work are given as follows:

$\circ$~\textit{Scheme 0: } Each 8×8 tile shares one spare row.

$\circ$~\textit{Scheme 1: } Two 8×8 tile shares two spare rows.

$\circ$~\textit{Scheme 2: } Two 8×8 tiles share three spare rows.

$\circ$~\textit{Scheme 3: } Three 8×8 tiles share three spare rows.

$\circ$~\textit{Scheme 4: } Three 8×8 tiles share four spare rows.

$\circ$~\textit{Scheme 5: } Four 8×8 tiles share four spare rows.

$\circ$~\textit{Scheme 6: } Four 8×8 tiles share five spare rows.

$\circ$~\textit{Scheme 7: } Five 8×8 tiles share four spare rows.

\section{Experimental Results and Analysis}
\label{sec:cache}
In this section, we firstly characterized the oscillation delay of a seven-stage RO array in a CNT-based FPGA. Then, the test configuration overhead of different scales of CNT-based FPGAs were evaluated. Moreover, we simulated the test application time for a single CNT-based CLB constructed by different input LUTs. In addition, the average test coverage, test overheads with different m-CNT distribution probabilities and initial jump sizes were evaluated. Finally, we evaluated the repair rate and hardware overheads under different spare row sharing architectures.

\subsection{Experimental Setup}
Process parameter settings of CNT in our experiments were the same as \cite{b20}. We generated a sample of CNT-based FPGA. Its structure is similar to Xilinx Virtex 7V2000T and consists of 391×391 FPGA tiles. For delay faults testing, the process variation of MWCNT was set according to \cite{b3}.

\begin{table}[!h]
	\centering
	\caption{~CNT Parameters}
	\vspace{0.1cm}
	\begin{tabular}{c}
		\includegraphics[width=1\linewidth]{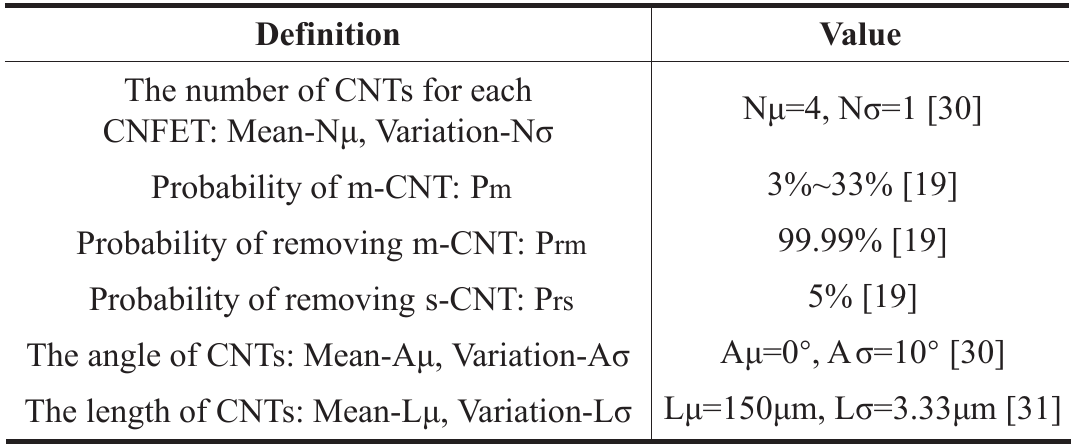}
	\end{tabular}
	
	\label{tab:par3}
	\vspace{-0.6cm}
\end{table}

To evaluate the effectiveness of our proposed CLB testing technique, we built a simulator with the layout information of the CNT-based FPGAs, and the parameter setting is shown in Table \ref{tab:par3}. We took the imperfect m-CNT removal process into consideration. The probability of m-CNTs is $p_{m}$, and $p_{Rm}$ is m-CNT removal rate. The starting coordinates of m-CNT were randomly generated. The CNT length had a mean length and standard deviation of $L_{\mu}$ and $L_{\sigma}$. The misaligned m-CNTs were randomly generated based on Gaussian distribution with misaligned probability as $p_{mis}$. According to \cite{b10}, for an architecture with the cluster size of 8, we estimated the footprint of a baseline CNT-based CLB to be 27698T where T denotes a minimum width transistor area, i.e. 2.2×$10^{-3}\mu{m^{2}}$ in 7nm technology node. Then, the area of a CLB can be estimated according to \cite{b20}. For a CLB composed of 4 six-input LUTs, m-CNTs in CNT bundles were randomly distributed based on the above parameter settings. We performed Monte-Carlo simulations to generate 1000 basic samples of the CNT-based FPGAs. After applying the recursive jump test with different jump steps and different misaligned angles of m-CNTs in these samples, we can derive the corresponding fault maps. Then, we compared the recursive jump testing with single-step and fixed-step testing schemes in terms of test coverage and test overhead. 
Finally, to evaluate the effectiveness of the redundant spare row sharing architecture, we evaluated the repair rate and hardware overhead for scheme 0$\sim$7. Two metrics used in comparisons are defined as follows:

Repair Ratio: The percentage of faulty segments repaired by redundant rows.

Spare segment overhead: The average number of redundant rows allocated per 8×8 tiles.

\subsubsection{The Delay Fault Testing of MWCNT Interconnects}
In this experiment, each RO had seven-stages and was placed in a CLB. ROs were mapped to the LUT by the XNOR configuration mentioned in Section \ref{sec:man}. Related work shows that it is feasible to measure the process variation or aging by mapping RO into FPGA \cite{b24}. In this work, the ROs were used to test the delay fault of MWCNT interconnects. To avoid the measurement noise, each frequency was measured three times and the average value was used. 

\begin{figure}[!t]
	\centering
	\includegraphics[width=1\linewidth]{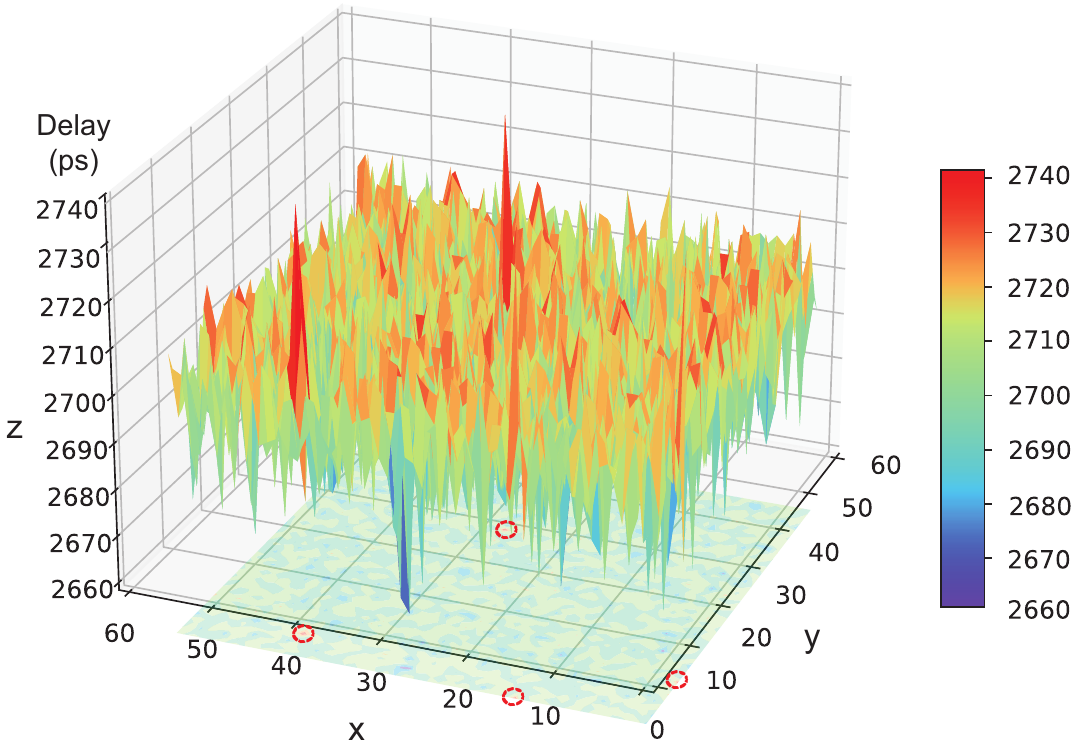}
	\vspace{-0.5cm}
	\caption{~The observed oscillation delay for each 7-stage oscillator in a CNT-based FPGA.}
	\label{fig:7ro56}
	\vspace{-0.3cm}
\end{figure}

Fig. \ref{fig:7ro56} shows the observed oscillation delay in one test configuration. The mean of oscillation delay was 2.70$ns$, and the variation was 100$ps$. Although the total range of variation is reasonably small, there are still a few ROs with large loop delays, which seriously affect the performance of a CNT-based FPGA operating at hundreds of MHz. So with ROs delay testing technique, the timing delay fault in MWCNT can be detected effectively.

These results demonstrate the suitability of RO-based testing for capturing the fine-grained delay variations unique to MWCNT interconnects, which traditional delay fault tests often overlook.

\subsubsection{Testing Overhead for m-CNT Faults in CLBs}
We evaluated the test overhead for different CLB array sizes, and the length of configuration bitstream and clock were set according to \cite{b14}.
\begin{figure}[!t]
	\centering
	\includegraphics[width=0.85\linewidth]{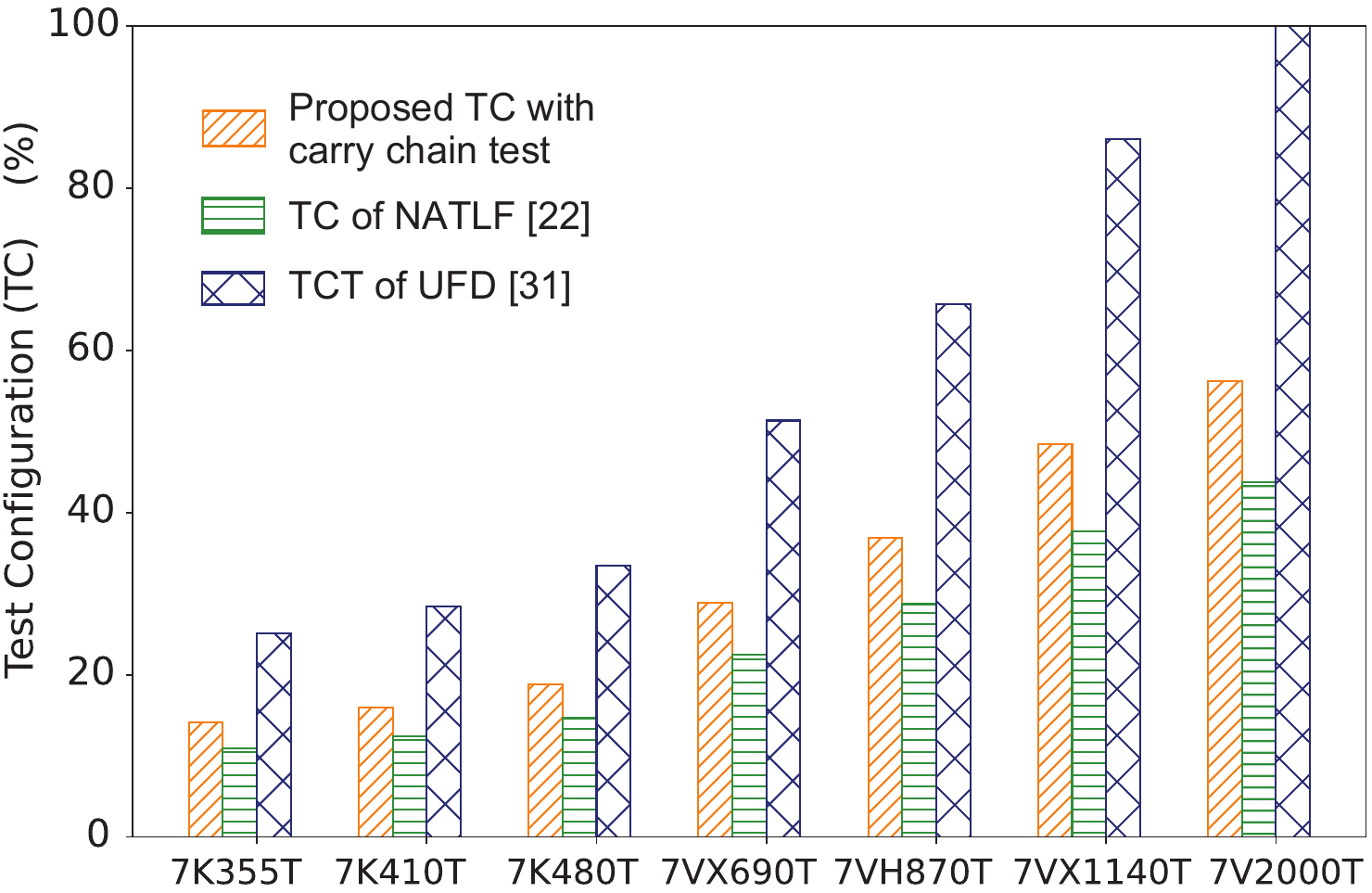}
	\vspace{-0cm}
	\caption{~Comparisons of the number of test sessions of different Xilinx-7 devices.}
	\label{fig:tc}
	\vspace{-0.2cm}
\end{figure}

Since the testing of carry chain was included, the test scheme proposed in this article requires $K+3$ test configurations on the basis of the traditional $K+1$ test configuration, where $K$ is the number of input ports in a LUT. As shown in Fig. \ref{fig:tc}, the test configuration overhead of the proposed technique (including the carry chain test) is slightly higher than the traditional NATLF method without carry chain test (increased by an average of 8.1\%) \cite{b18}, but far less than the traditional UFD method \cite{b27}. In summary, adding the test of carry chain in the CLB only incurs little test overhead.
\begin{figure}[!t]
	\centering
	\includegraphics[width=0.85\linewidth]{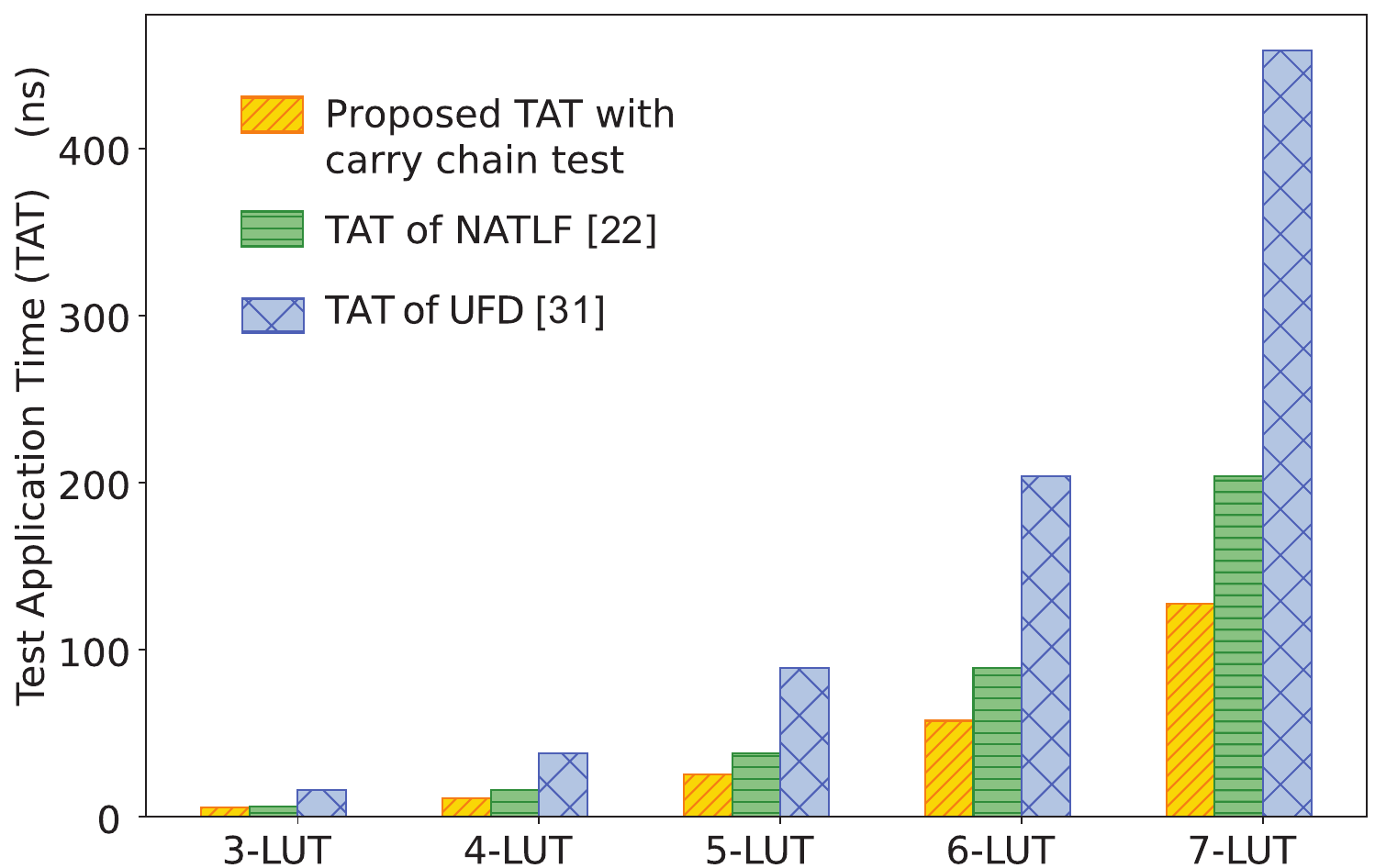}
	\vspace{-0cm}
	\caption{~Simulation results for the test time of different CNT-based LUTs.}
	\label{fig:ta}
	\vspace{-0.6cm}
\end{figure}

Then, we applied the technique mentioned in Section \ref{sec:part} to a single CNT-based CLB constructed by different input LUTs. We evaluated the test time in one test session by SPICE simulation. As shown in Fig. \ref{fig:ta}, the test time (including the carry chain test) of the proposed technique is less than the other two traditional methods without carry chain test (i.e., NATLF \cite{b18}, UFD \cite{b27}), and the test time decreases more significantly with the increase of LUT input ports. Compared with the NATLF \cite{b18}, the test time decreased by 28.77\% on average. For the general 6-input LUT, the test time can be reduced by 35.49\%.

The proposed RO-based test with carry chain addition requires only marginally more configurations than NATLF, but achieves more comprehensive fault coverage. Therefore, it offers a better coverage-to-cost trade-off than both NATLF and UFD.

\subsection{Experimental Results and Discussion}

\subsubsection{Evaluations of Cascaded Faulty CLB Segment Testing}
There are two major methods for FPGA testing: application-independent testing and application-dependent testing. For application-independent testing, since the actual user configuration is unknown during testing, so all FPGA logic units will be tested. The application-dependent testing only partially tests the circuit actually used \cite{b28}.

\begin{table}[ht]
\centering
\caption{Comparison of test methods}
\renewcommand{\arraystretch}{1.2}

\begin{tabular}{|>{\centering\arraybackslash}p{1.2cm}|
                >{\centering\arraybackslash}p{2.8cm}|
                >{\centering\arraybackslash}p{2.9cm}|}
\hline
\textbf{Method} & \textbf{Fault Coverage} & \textbf{Test Overhead} \\
\hline
Single & Baseline (100\%) & Baseline (100\%) \\
Fixed  & Mid (45.4–78.2\%) & Low (2.9–31.55\%) \\
Recursive & High (82.9–100\%) & Mid (64.2–69.64\%) \\
\hline
\end{tabular}

\vspace{1mm}
\begin{flushleft}
\small
\vspace{-0.3cm}
\textbf{Note: Recursive test starts with jump = 4.}
\end{flushleft}
\label{tab:test-methods}
\end{table}



\begin{figure}[!t]
\vspace{-0.6cm}
	\centering
	\subfloat[]{
		\includegraphics[width=1.6in]{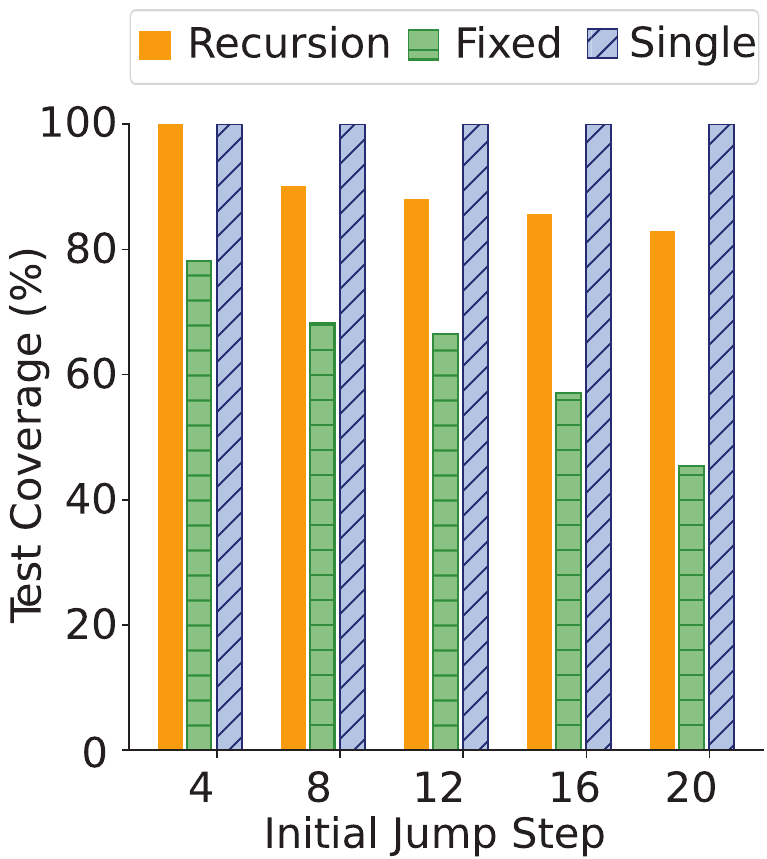}}
	\subfloat[]{
		\includegraphics[width=1.6in]{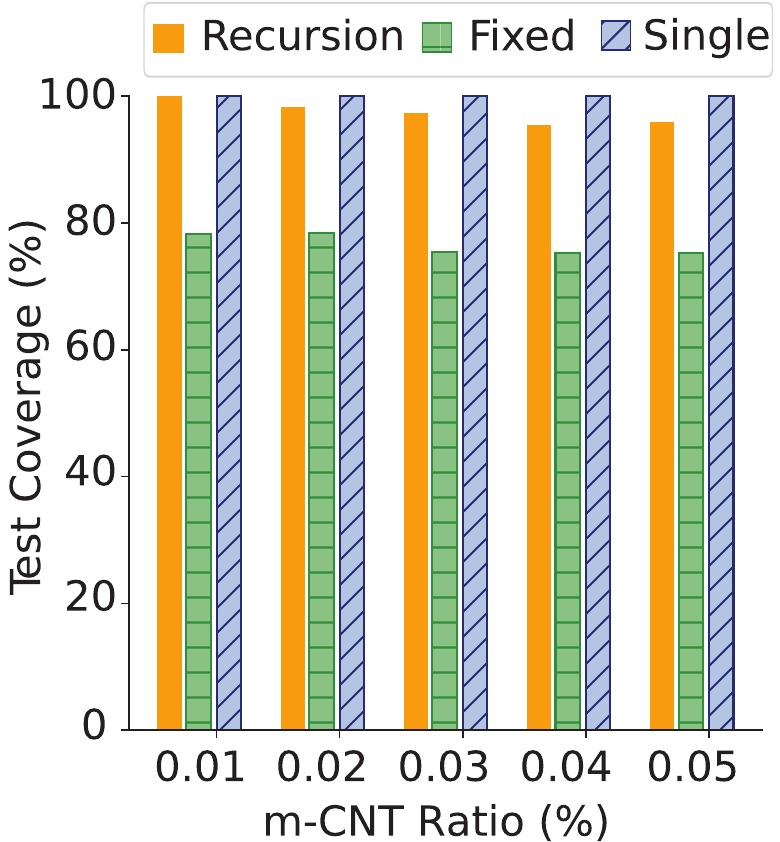}}
	\caption{~~ Simulation results on test converage with varying (a) jump step, (b) m-CNT ratio.} 
	\label{fig:buco}
\end{figure}

For application-independent testing, we compared the recursive testing with the fixed-step jump testing and single-step testing. The m-CNT probability was assumed to be 0.01\%, and the default size of initial jump step size was set to 4. We evaluated the average test coverage obtained by varying initial jump step size and m-CNT ratio. As shown in Fig. \ref{fig:buco}(a), the proposed jump tests with step size 4 show 100\% test coverage. The fixed-step jump testing, however, results in lower test coverage as the initial jump step increases. This is because it has a higher chance to jump over the start/end point of the faulty CLB segment. As shown in Fig. \ref{fig:buco}(b), the recursive test provides higher test coverage than the fixed-step jump test as m-CNT probability increases. The average test coverage of recursive test is 96.58\%, which is much higher than fixed-step jump test (63.80\%). Therefore, the recursive jump method strikes the best balance.

\begin{figure}[!t]
	\vspace{-0.6cm}
	\centering
	\subfloat[]{
		\includegraphics[width=1.5in]{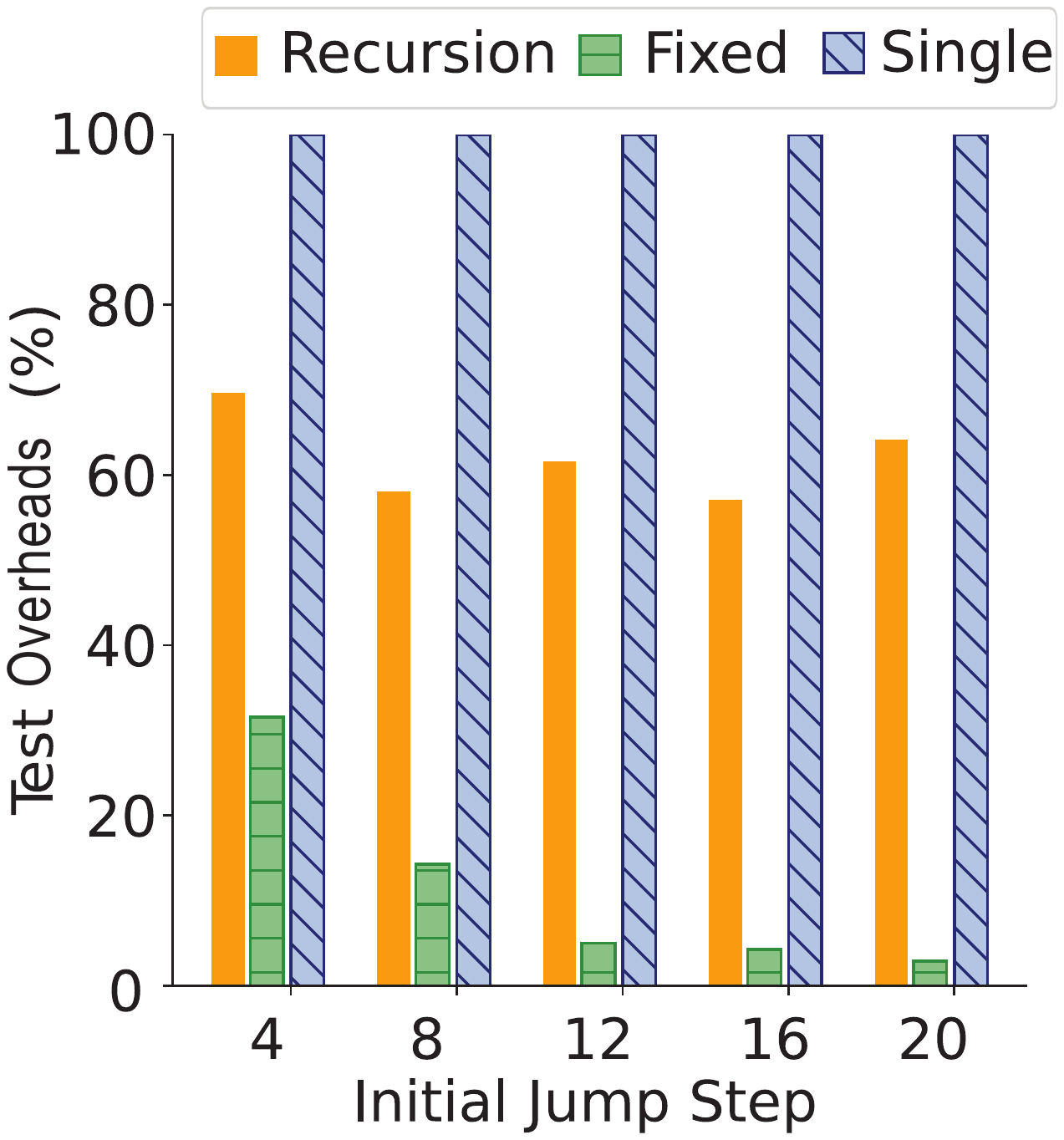}
	}
	\subfloat[]{
		\includegraphics[width=1.5in]{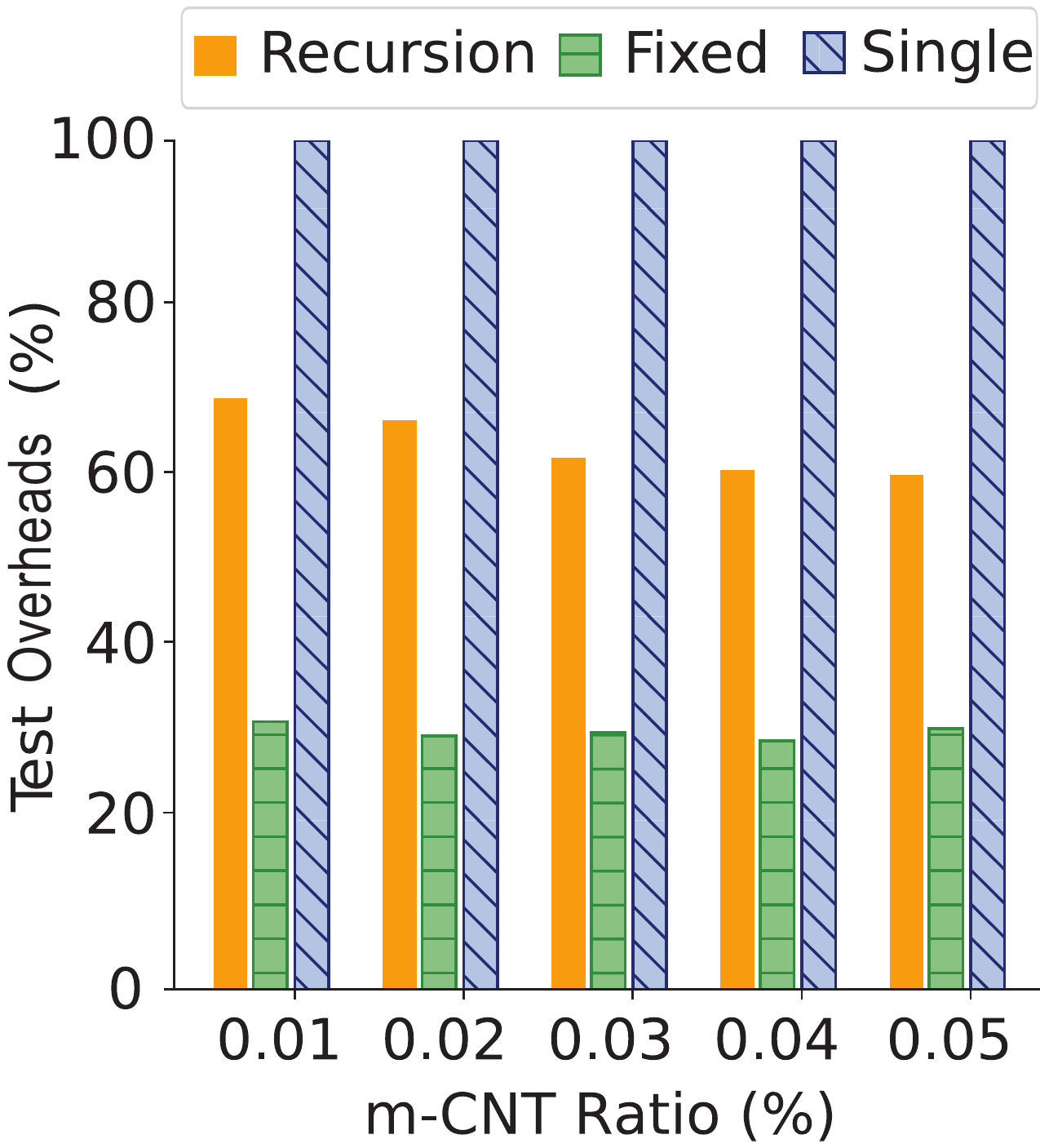}
	}
	\quad
	\vspace{-0.2cm}
	\caption{~~ Simulation results on test overheads with varying (a) jump step, (b) m-CNT ratio.}
	\label{fig:mco}
    \vspace{-0.6cm}
\end{figure}

Fig. \ref{fig:mco} presents the test overheads, which are normalized to the overheads of single-step tests. As expected, the test overhead of the recursive test is higher than that of the fixed-step jump test. This is because more jumps are required to guarantee high fault coverage, and the test overhead of the recursive testing slightly increases with the increase of m-CNT probability. Compared with the single-step testing, the test overhead of recursive testing can be reduced by 35.78\% on average.

In particular, Fig. \ref{fig:mco}(a) reveals an interesting phenomenon: the test overhead with jump step size 12 is higher than that with jump step size 8. This is because when the recursive step is reduced to 3 or 5, the algorithm increases the last jump step size by 1 and divides it by 2. For example, for the initial jump step 8, the recursive jump test would use jump steps 8, 4, 2, and 1, while for the initial jump step size 12, recursive jump test would use jump steps 12, 6, 3, 2 and 1, which brings higher test overheads. Similar results can be observed with the initial jump step size 20. Based on this observation, we suggest set the initial test step size as four. Experimental results show that the proposed recursive testing can achieve a high test coverage with low test overhead.

For application-dependent test, we also used the recursive jump test method, and the initial jump step size is four and the m-CNT probability is 0.05\%. At the beginning, we chose 10 different benchmarks and imported them into Vivado. Then the benchmarks were synthesized to generate gate netlists, and programmable logic blocks used in the FPGA array were identified. We used recursive testing, fixed-step testing and single-step testing, respectively. Test coverages and test overheads with different benchmarks are shown in Table \ref{tab:par4}. We can observe that the test coverage of the recursive testing is higher than the fixed-step jump test while the test overhead of the recursive testing is lower than the single-step testing, which validates the effectiveness of our proposed recursive testing technique.

\begin{table} [!h]
	\vspace{-0.2cm}
	\centering
	\caption{~Application-dependent test}
	\vspace{0.1cm} 
	\setlength{\tabcolsep}{0.6mm}{
		\begin{tabular}{*{7}{c}}
			\toprule
			\multirow{2}*{Benchmark} & \multicolumn{3}{c}{Test Coverage (\%)} & \multicolumn{3}{c}{Test Overhead (\%)} \\
			\cmidrule(lr){2-4}\cmidrule(lr){5-7}
			& Recursion & Fixed & Single & Recursion & Fixed & Single \\
			\midrule
			PCI & 91 & 71 & 100 & 33 & 19 & 100 \\
			I2C & 84 & 63 & 100 & 31 & 16 & 100 \\
			SPI & 93 & 69 & 100 & 47 & 23 & 100 \\
			FIR & 79 & 60 & 100 & 36 & 27 & 100 \\
			FPU & 82 & 66 & 100 & 44 & 26 & 100 \\
			VGA & 92 & 70 & 100 & 49 & 29 & 100 \\
			PCM & 81 & 73 & 100 & 61 & 21 & 100 \\
			DMA & 87 & 59 & 100 & 54 & 19 & 100 \\
			USB & 82 & 64 & 100 & 32 & 15 & 100 \\
			MEM & 86 & 61 & 100 & 29 & 16 & 100 \\
			\bottomrule
			\label{tab:par4}
			\vspace{-0.2cm}
		\end{tabular}
		\label{tab:par}
		\vspace{-0.6cm}
	}
\end{table}

As summarized in Table~\ref{tab:test-methods}, the recursive jump test consistently achieves high coverage (82.9–100\%) with moderate overhead. This observation is further validated in Table~\ref{tab:par}, demonstrating recursive jump practicality in real-world application scenarios.

\subsubsection{Testing Overhead and Effectiveness of Redundant Sharing Architecture}
\begin{figure}[t]
\vspace{-0cm}
	\centering
	\includegraphics[width=0.99\linewidth]{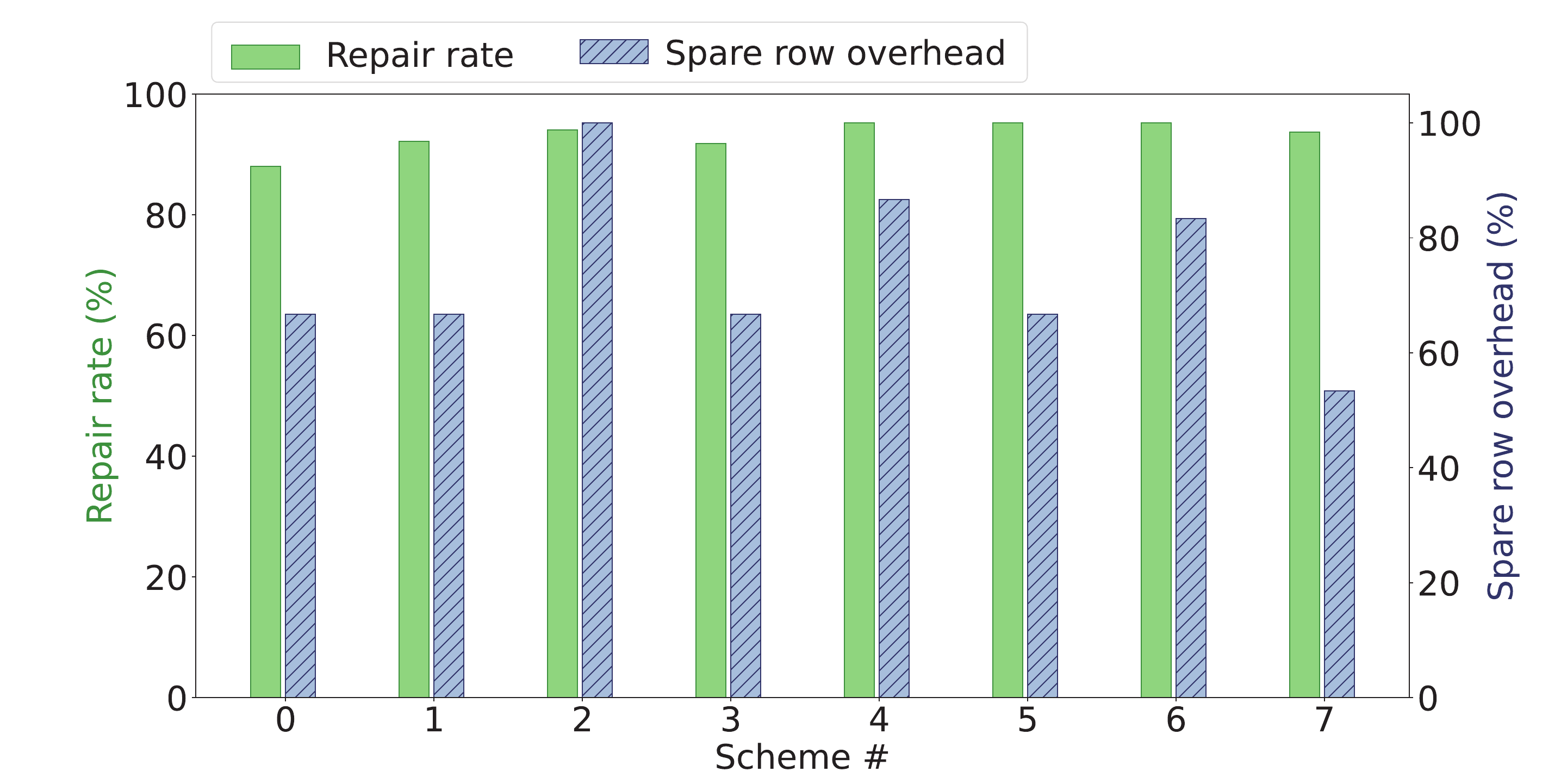}
	\vspace{-0.2cm}
	\caption{~Repair rate and spare row overhead for redundancy architecture.}
	\label{fig:rep}
	\vspace{-0.6cm}
\end{figure}

Fig. \ref{fig:rep} shows the repair rate and hardware overhead of the proposed spare row sharing architecture with different sharing strategies. From the experimental results, we can see that scheme 5: four 8×8 tile sharing four spare rows, has the highest repair rate and less hardware overhead than most of other schemes.

\begin{table}[t]
\centering
\caption{Comparison of spare row sharing schemes} 
\renewcommand{\arraystretch}{1.2}

\begin{tabular}{|>{\centering\arraybackslash}m{1.3cm}|
                >{\centering\arraybackslash}m{2.3cm}|
                >{\centering\arraybackslash}m{2.3cm}|}
\hline
\textbf{Scheme} & \textbf{Repair Rate} & \textbf{Overhead} \\
\hline
Scheme 0 & Low(92.4\%) & Mid(66.7\%)  \\
\hline
Scheme 2 & High(98.8\%) & Baseline(100\%)  \\
\hline
Scheme 5 & High(100\%) & Mid(66.7\%) \\
\hline
Scheme 7 & High(98.4\%) & Low(53.3\%)\\
\hline
\end{tabular}

\label{tab:spare-schemes}
\end{table}


\vspace{0.5cm}

As the number of sharing units increases, circuit area, latency, and power consumption also increase significantly \cite{b12.4}, so there should not be too many sharing units. Looking at scheme 7: five 8×8 tiles sharing four spare rows. The repair rate loses 1.2\%, which is negligible compared with scheme 5. However, the hardware overhead of scheme 7 is significantly reduced by 13.34\%$\sim$46.67\% compared with other schemes.  

As shown in Table~\ref{tab:spare-schemes}, Scheme~7 achieves a high repair rate (98.4\%) comparable to the best-performing configuration, while reducing hardware redundancy by over 46.7\%. This makes it the most efficient and scalable solution among the evaluated schemes.

\subsubsection{ Fault Injection and Detection Coverage Evaluation}

To validate the fault detection capability of the proposed testing framework, we conducted a logic-level fault injection experiment on CNT-based FPGA tile.
\begin{figure}[!h]
    \centering
    \includegraphics[width=0.95\linewidth]{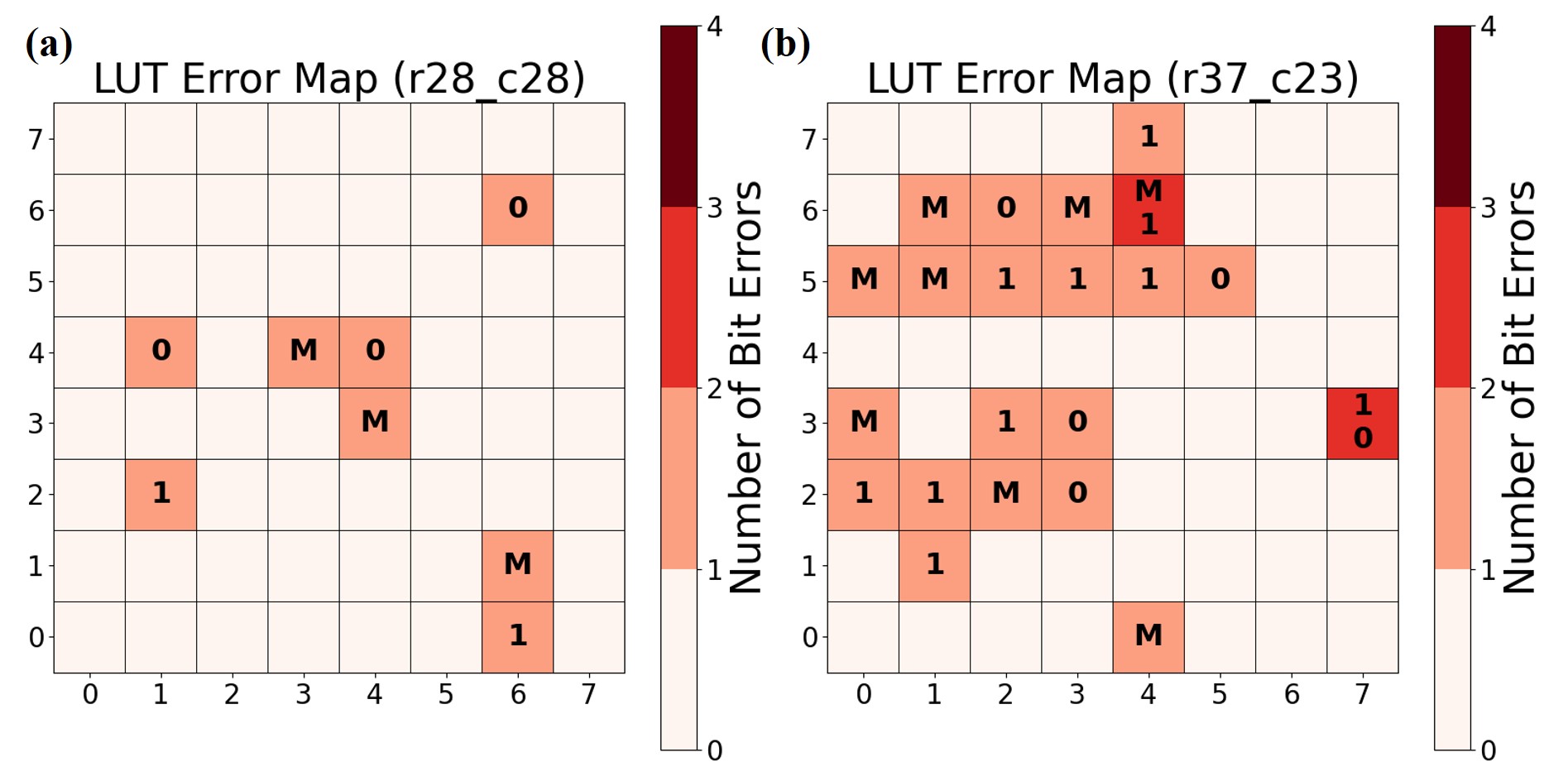}
    \vspace{-0.2cm}
    \caption{Visualization of injected logic faults in LUTs. 
    (a) Randomly distributed fault pattern involving all three fault types.
    (b) Clustered fault region reflecting correlated CNT-induced defects.}
     \vspace{-0.2cm}
    \begin{flushleft}
        \scriptsize Note: “0” indicates stuck-at-0 fault; “1” indicates stuck-at-1 fault; “M” indicates mux override fault.
    \end{flushleft}
    \label{fig:fault_inject_maps}
    \vspace{-0.2cm}
\end{figure}



Faults were injected directly at the LUT level, where each 6-input LUT was assigned one of three representative fault models:

$\circ$~\textit{stuck-at-0}: The LUT output is permanently forced to logic 0, regardless of input selection.

$\circ$~\textit{stuck-at-1}: The LUT output is permanently forced to logic 1, regardless of input selection.

$\circ$~\textit{mux override}: The internal address decoding logic is overridden, causing the output to be fixed to a specific configuration bit regardless of the intended \textit{sel} input.

Fig.~\ref{fig:fault_inject_maps} presents the spatial distribution of injected faults in a single 8$\times$8 tile block. Each tile contains four LUTs, and the injected faults are marked according to their type. In Fig.~\ref{fig:fault_inject_maps}(a), faults of all three types appear scattered across the block, illustrating the diversity and randomness of possible CNT-induced logic errors. In contrast, Fig.~\ref{fig:fault_inject_maps}(b) exhibits a correlated defect region where adjacent tiles exhibit clusters of faults, which is consistent with physical CNT alignment anomalies and bundle-induced disruptions.

For fault detection, we adopted the recursive jump testing strategy at the \textit{tile level}. That is, test patterns were generated and applied to tiles as atomic units. Internally, to accelerate fault localization within a tile, we leveraged the LUT-level configuration-aware structure introduced in Section~\ref{5_1}. This structure enables fast lookup of erroneous LUTs by checking aggregated tile responses.
The detection process was applied across all 49$\times$49 simulated blocks, each consisting of 8$\times$8 tiles, thereby covering the full 391$\times$391 tile array. A fault was considered detected if any of the four LUTs within a tile produced an output mismatch under recursive test stimuli. To evaluate effectiveness, we compared the detection performance of the recursive, fixed-step, and ideal single-step test strategies across the entire array.

\begin{table}[h]
    \vspace{-0.2cm}
    \centering
    \caption{~Detected Fault Count and Coverage by Method}
    \vspace{0.1cm} 
    \setlength{\tabcolsep}{1.2mm}{
    \begin{tabular}{c c c c}
        \toprule
        \textbf{Fault Type} & \textbf{\textbf{Recursion}} & \textbf{Fixed} & \textbf{Single} \\
        \midrule
        MUX        & \textbf{13126 } & 8634   & 14693  \\
        Stuck-at-0 & \textbf{19651 } & 13030  & 22121  \\
        Stuck-at-1 & \textbf{13115 } & 8843   & 14764  \\
        \midrule
        \textbf{Total} & \textbf{45892 (89.0\%)} & 30407 (58.9\%) & 51578 (100\%) \\
        \bottomrule
    \end{tabular}
    \label{tab:inject_fault_compact}
    \vspace{-0.4cm}
    }
\end{table}

As summarized in Table~\ref{tab:inject_fault_compact}, the proposed recursive method detects over 88\% of injected LUT faults across all types, significantly outperforming fixed-step testing and closely approaching the ideal fault detection coverage achieved by exhaustive single-step tests.

\color{black}

\section{Conclusion}
\label{sec:con}
CNT-based FPGA is a promising alternative to conventional CMOS-based FPGA. However, due to the imperfect fabrication process of CNTs, CNT-based FPGA may exhibit unique faulty patterns. With the help of an advanced ring oscillator design, CLBs can be connected in series to form a ring oscillator, which can be used to effectively detect the delay fault of MWCNT interconnects. Furthermore, for the faulty CLBs induced by m-CNTs, we propose a carry chain test method based on the traditional test method, and a new technique is proposed to speed up the fault testing. Finally, considering the faulty segments induced by m-CNTs, we propose a fault tolerant architecture by sharing the spare rows to repair the faulty segments, which can improve the repair rate and reduce the hardware overhead effectively.

\section*{Acknowledgment}

The authors would like to thank Cheng Liu and Ying Wang from Institute of Computing Technology, Chinese Academy of Sciences, for providing many constructive suggestions during the development
of the motivation.

This work is supported in part by Shenzhen Science and Technology Program under Grant No. SGDX2023011609\linebreak3303006 and KJZD20231023100201003.

\end{document}